\documentclass[preprint2,longabstract]{aastex}

\newcommand{\beq}{\begin{equation}}
\newcommand{\eeq}{\end{equation}}
\newcommand{\bdi}{\begin{displaymath}}
\newcommand{\edi}{\end{displaymath}}
\newcommand{\snr}{S/N}
\newcommand{\kms}{km\,s$^{-1}$}

\slugcomment{To appear in the Astrophysical Journal}

\shorttitle{BLAST Survey in Vulpecula}
\shortauthors{Chapin, E.~L.~et al.}

\begin{document}

\title{ The Balloon-borne Large Aperture Submillimeter Telescope (BLAST) 2005:
  A 4~deg$^2$ Galactic Plane Survey in Vulpecula ($\ell=59^\circ$)
}

\author{E.~L.~Chapin,\altaffilmark{1,\dag}
        P.~A.~R.~Ade,\altaffilmark{2}
        J.~J.~Bock,\altaffilmark{3,4}
        C.~Brunt,\altaffilmark{5}
        M.~J.~Devlin,\altaffilmark{6}
        S.~Dicker,\altaffilmark{6}
        M.~Griffin,\altaffilmark{2}
        J.~O.~Gundersen,\altaffilmark{7}
        M.~Halpern,\altaffilmark{1}
        P.~C.~Hargrave,\altaffilmark{2}
        D.~H.~Hughes,\altaffilmark{8}
        J.~Klein,\altaffilmark{6}
        G.~Marsden,\altaffilmark{1}
        P.~G.~Martin,\altaffilmark{9,10}
        P.~Mauskopf,\altaffilmark{2}
        C.~B.~Netterfield,\altaffilmark{10,11}
        L.~Olmi,\altaffilmark{12,13}
        E.~Pascale,\altaffilmark{11}
        G.~Patanchon,\altaffilmark{1,14}
        M.~Rex,\altaffilmark{6}
        D.~Scott,\altaffilmark{1}
        C.~Semisch,\altaffilmark{6}
        M.~D.~P.~Truch,\altaffilmark{15}
        C.~Tucker,\altaffilmark{2}
        G.~S.~Tucker,\altaffilmark{15}
        M.~P.~Viero,\altaffilmark{10}
        D.~V.~Wiebe\altaffilmark{11}}

\altaffiltext{1}{Department of Physics \& Astronomy, University of
British Columbia, 6224 Agricultural Road, Vancouver, BC V6T~1Z1,
Canada}

\altaffiltext{2}{Department of Physics \& Astronomy, Cardiff University, 5 The Parade, Cardiff, CF24~3AA, UK}

\altaffiltext{3}{Jet Propulsion Laboratory, Pasadena, CA 91109-8099}

\altaffiltext{4}{Observational Cosmology, MS 59-33, California Institute of Technology, Pasadena, CA 91125}

\altaffiltext{5}{School of Physics, University of Exeter, Stocker
Road, Exeter, EX4~4QL, UK}

\altaffiltext{6}{Department of Physics and Astronomy, University of Pennsylvania, 209 South 33rd Street, Philadelphia, PA 19104}

\altaffiltext{7}{Department of Physics, University of Miami, 1320 Campo Sano Drive, Carol Gables, FL 33146}

\altaffiltext{8}{Instituto Nacional de Astrof\'isica \'Optica y Electr\'onica (INAOE), Aptdo. Postal 51 y 72000 Puebla, Mexico}

\altaffiltext{9}{Canadian Institute for Theoretical Astrophysics, University of Toronto, 60 St. George Street, Toronto, ON M5S~3H8, Canada}

\altaffiltext{10}{Department of Astronomy \& Astrophysics, University of Toronto, 50 St. George Street, Toronto, ON  M5S~3H4, Canada}

\altaffiltext{11}{Department of Physics, University of Toronto, 60 St. George Street, Toronto, ON M5S~1A7, Canada}

\altaffiltext{12}{Istituto di Radioastronomia, Largo E. Fermi 5, I-50125, Firenze, Italy}

\altaffiltext{13}{University of Puerto Rico, Rio Piedras Campus, Physics Dept., Box 23343, UPR station, San Juan, Puerto Rico}

\altaffiltext{14}{Laboratoire APC, 10, rue Alice Domon et L{\'e}onie Duquet 75205 Paris, France}

\altaffiltext{15}{Department of Physics, Brown University, 182 Hope Street, Providence, RI 02912}

\altaffiltext{\dag}{\url{echapin@phas.ubc.ca}}

\begin{abstract}
  We present the first results from a new 250, 350, and 500\,\micron\
  Galactic Plane survey taken with the Balloon-borne Large-Aperture
  Submillimeter Telescope (BLAST) in 2005. This survey's primary goal
  is to identify and characterize high-mass proto-stellar objects
  (HMPOs).  The region studied here covers 4\,deg$^2$ near the open
  cluster NGC~6823 in the constellation Vulpecula ($\ell=59^\circ$).
  We find 60 compact sources ($<60''$ diameter) detected
  simultaneously in all three bands.  Their spectral energy
  distributions (SEDs) are constrained through BLAST, {\it IRAS}, {\it
    Spitzer} MIPS, and {\it MSX} photometry, with inferred dust
  temperatures spanning $\sim 12$--40\,K assuming a dust emissivity
  index $\beta=1.5$.  The luminosity-to-mass ratio, a
  distance-independent quantity, spans
  $\sim0.2$--130\,L$_\odot$\,M$_\odot^{-1}$. Distances are estimated
  from coincident $^{13}$CO$(1 \rightarrow 0)$ velocities combined
  with a variety of other velocity and morphological data in the
  literature. In total, 49 sources are associated with a molecular
  cloud complex encompassing NGC~6823 (distance $\sim 2.3$\,kpc), 10
  objects with the Perseus Arm ($\sim 8.5$\,kpc) and one object is
  probably in the outer Galaxy ($\sim 14$\,kpc).  Near NGC~6823, the
  inferred luminosities and masses of BLAST sources span $\sim
  40$--$10^4$\,L$_\odot$, and $\sim15$--$700$\,M$_\odot,$
  respectively.  The mass spectrum is compatible with molecular gas
  masses in other high-mass star forming regions.  Several luminous
  sources appear to be Ultra Compact \ion{H}{2} regions powered by
  early B stars. However, many of the objects are cool, massive
  gravitationally-bound clumps with no obvious internal radiation from
  a protostar, and hence excellent HMPO candidates.
\end{abstract}


\section{INTRODUCTION}     
\label{sec:intro}

The Balloon-borne Large Aperture Submillimeter Telescope (BLAST) is a
2-m stratospheric balloon telescope that observes simultaneously at
250, 350, and 500\,\micron\ using bolometric imaging arrays
\citep{pascale2007}.  During the first BLAST science flight, a 100~hr
Arctic flight from Sweden to Canada in June 2005 (BLAST05), BLAST
conducted the first sensitive large-scale Galactic Plane surveys at
these wavelengths.  The focus of these surveys is the earliest stages
of massive star formation. As noted in the recent review by
\citet{zinn2007} our understanding of this evolutionary phase is still
limited.  Over the past 15 years submillimeter observations at longer
wavelengths (350--1200\,\micron), e.g. with SCUBA on the 15-m James
Clerk Maxwell Telescope \citep{holland1999}, or MAMBO on the IRAM 30-m
telescope \citep{kreysa1998,motte2007}, have opened up studies of the
earliest evolutionary stages of molecular core collapse and
proto-stellar formation.  Such cores lack an internal source of
radiation and are very cold ($\la 25$\,K), so that they emit the bulk
of their radiation at submillimeter wavelengths. Since this emission
is optically thin in the submillimeter band, observed flux densities
are proportional to column density and mass.  There is a variety of
terminology used for these elusive early stages.  For example,
\citet{zinn2007} distinguish cores (size 0.1\,pc) embedded within
clumps (size 0.5\,pc).  \citet{motte2007} refer to molecular cloud
fragments which could be high-luminosity infra-red protostars,
infra-red quiet protostars, or high-mass pre-stellar cores. Along the
lines of the latter, here we refer to high-mass proto-stellar objects
(HMPOs) as compact sources residing in dense molecular clouds, that
have the potential to form (one or more) massive OB stars, having
luminosities in the range $\sim10^2$--$10^5$\,L$_{\odot}$, but without
associated radio continuum emission.  This latter qualification
distinguishes them from massive young stellar objects, a later stage
in which a hot star has formed (perhaps still accreting), providing
the ionizing radiation necessary to form a high emission measure Ultra
Compact (UC) \ion{H}{2} region (and hence radio emission).  Note that
at the distances of high-mass star forming regions surveying
instruments detect quite massive clumps, possibly harboring groups of
stars or their precursors, and not resolving the substructure
associated with individual nodes of collapse; however, once star
formation is underway the most massive objects dominate the luminosity
and ionization.

BLAST is presently unique in its ability to detect and characterize
cold dust emission from a range of pre- and proto-stellar sources,
constraining the temperatures of objects with $T \la 25$\,K
($\beta=1.5$) using its three-band photometry near the peak of the
spectrum.  An earlier, less sensitive balloon-borne submillimeter
telescope called ProNaOS \citep{dupac2001} had similar goals.
Previous studies which do not sample the spectral peak of the thermal
emission \citep[e.g.,][]{johnstone2000, pierceprice2000, kirk2005,
reid2005, thompson2005, enoch2006, hill2006, schneider2006,
thompson2006, young2006, enoch2007, moore2007} have been limited by
their relative inability to measure the temperature, producing large
uncertainties in the derived luminosities and masses.  Recent surveys
with {\it Spitzer} MIPS can constrain the temperatures of warmer
objects \citep{car05}, but the youngest and coldest objects are
potentially not detected even in the long-wavelength {\it Spitzer}
bands.

\begin{figure*}
\centering
\includegraphics[width=\linewidth]{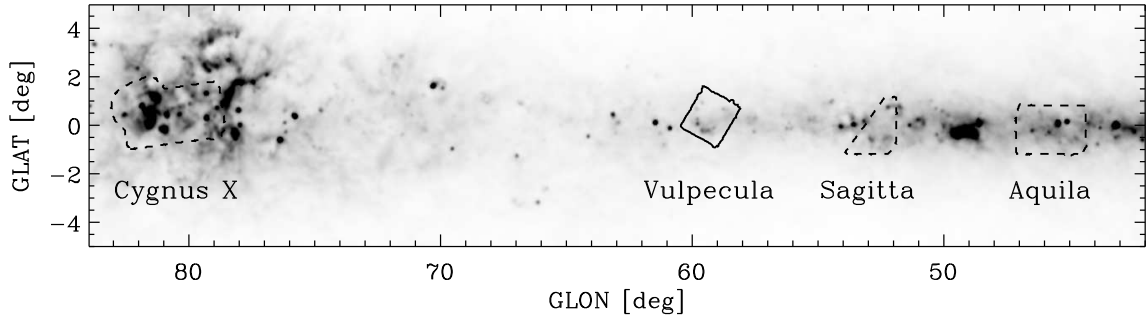}
\caption{Locations of fields for the BLAST05 Galactic Plane surveys.
  Results from the Vulpecula field are discussed in detail in this
  paper. The background intensity image is 100\,\micron\ emission from
  \citet{schlegel1998}.}
\label{fig:coverage}
\end{figure*}

BLAST is also very efficient at mapping, and so conducted a series of
surveys spread across the entire portion of the Galactic Plane that
was available during the flight.  These maps encompass (primarily)
known high-mass star forming regions: a 10\,deg$^2$ map of Cygnus X; a
4\,deg$^2$ map in Vulpecula (described in detail in this paper); a
3\,deg$^2$ map in Sagitta; and finally a 6\,deg$^2$ map in Aquila
towards the Galactic Ring Survey molecular cloud GRSMC~45.60+0.30
\citep{rathborne2004}. The locations and sizes of these maps are shown
in Figure~\ref{fig:coverage}. The Vulpecula map was studied first, as
it is the deepest of the four regions, and consists of 6.5 hours of
data. The analysis of the other three fields, and those from the 2006
Antarctic flight, including a diffraction-limited 50\,deg$^2$ map of
the Vela molecular ridge, will be presented in forthcoming papers.

To put these surveys in context, \citet{motte2007} state that ``large
fast-mapping images with MAMBO-2 are currently the best tool to study
density structure of molecular clouds with high spatial resolution.''
In about 33 hours of observations (and 30 more with the less efficient
MAMBO) they mapped 3\,deg$^2$ in Cygnus~X; not the entire region but
targeted areas of high extinction.  They catalogued 129 compact
sources (down to a scale of 15\arcsec) to a peak flux level of 80\,mJy
($5\,\sigma$) and proposed an additional 40 somewhat more extended
structures (2\arcmin).  Larger structures (10\arcmin) were filtered
out.
In the survey reported here, we mapped a contiguous (no threshold for
extinction) 4\,deg$^2$ in 6.5 hours.  Although the very largest scales
(including the DC level) are absent from our maps, our images maintain
power well beyond 10\arcmin; the images reveal interesting large-scale
structures crossing the entire region.  As reported below, we catalog
60 sources to a flux level of 10\,Jy at 250\,\micron.  Extrapolating
this detection limit using a typical spectral shape ($T = 20$\,K and
$\beta = 1.5$) this corresponds to 170\,mJy at 1.2\,mm.  We also find
fewer sources per square degree compared to the \citet{motte2007}
survey of Cygnus~X because this region is slightly more distant, not
as active, and our survey did not target areas of high extinction.
There are exciting large sensitive surveys to look forward to at these
same wavelengths with the SPIRE instrument for {\it Herschel}
\citep{griffin2003}.

The region chosen for this study was centered near NGC~6823 in the
constellation Vulpecula ($\ell=59^\circ$), and covers approximately
4\,deg$^2$. This region of the Galactic Plane is prominent in images
of thermal dust emission (e.g., {\it IRAS} in
Figure~\ref{fig:coverage}), as well as in the radio and the optical.
In the radio, the identifiable extended \ion{H}{2} region excited by
the massive stars is called Sh2-86 (nominal position indicated in
Figure~\ref{fig:blastmaps}).  The stellar HR diagram for NGC~6823 has
been examined by \citet{massey1995}.  They find an age of 5--7\,Myr
for the bulk of the stars.  There are several evolved OB supergiants
of mass $\sim 25$\,M$_\odot$.  The most massive star, O7 V ((f)), with
mass $\sim40$\,M$_\odot$, appears younger (2\,Myr) than the rest.  In
the optical this massive star is responsible for the illumination of
many ``elephant trunks'' or ``pillars'' in the eastern portion of the
\ion{H}{2} region.

Very high-mass stars evolve quickly into supernovae.  We note that a
supernova remnant (SNR) discovered by \citet{taylor1992} is within
20\arcmin\ of the O7 star in projection ($< 12$\,pc laterally,
assuming the same distance), and even closer to one B supergiant ---
nominal positions are indicated in Figure~\ref{fig:blastmaps}.
\citet{massey1995} find several pre-main sequence stars of mass
5--7\,M$_\odot$, which pass through this stage in less than 2\,Myr.
Earlier stages of evolution are revealed by the bright Infra Red
Astronomy Satellite ({\it IRAS}) sources clustered in this region.  At
least seven of these objects have been observed in molecular line
studies \citep{beu02,zha05,bel06}

This paper describes a technique developed for deconvolving the
out-of-focus beams common to all data from BLAST05
\citep[see][]{truch2007} and for detecting compact objects in the
BLAST maps. We report a robust list of 60 submillimeter sources with
sizes $\la60''$ diameter that are detected simultaneously in the three
BLAST bands (\S\ref{sec:obs}).  Infra-red photometry for BLAST sources
is obtained from comparisons with {\it IRAS} and {\it Spitzer} MIPS
maps, and the Midcourse Space Experiment ({\it MSX}) point source
catalog (\S\ref{sec:firsed}).  The BLAST and infra-red data are
combined to constrain dust temperatures and integrated far infra-red
(FIR) fluxes using isothermal modified blackbody SED fits
(\S\ref{sec:coldsed}).  Our analysis has benefited from a $^{13}$CO$(1
\rightarrow 0)$ data cube obtained as a part of a Galactic Plane
survey at FCRAO \citep{brunt2007}. Noting coincidences between compact
structures in the $^{13}$CO emission, and comparing velocities and
morphologies with the VLA Galactic Plane Survey, {\it Spitzer} GLIMPSE
\citep{whitney2005} and Digitized Sky Survey images, distance
estimates are obtained for all of the sources in \S\ref{sec:dist}. Of
60 BLAST sources, 49 objects are likely associated with the molecular
clouds surrounding NGC~6823 at $\sim2.3$\,kpc, 10 appear to be in the
more distant Perseus arm ($\sim8.5$\,kpc), and a single object is
thought to lie in the outer Galaxy ($\sim14$\,kpc). These distances
are used to calculate source luminosities and masses in
\S\ref{sec:mass}.  In this section we also present an estimate of the
mass function for sources associated with NGC~6823.  In this survey
high-mass clumps are seen in a range of evolutionary stages: those
with embedded UC \ion{H}{2} regions; luminous objects without
substantial radio emission (HMPOs); and cold gravitationally-bound
clumps (or cores) with low luminosity-to-mass ratio and no evidence of
star formation (\S\ref{sec:disc}).

\section{BLAST OBSERVATIONS TOWARDS NGC~6823} 
\label{sec:obs}

\subsection{Observing Strategy}

BLAST05 is a 2-m Cassegrain telescope, whose under-illuminated primary
mirror is designed to produce diffraction-limited beams with FWHM
40\arcsec, 58\arcsec, and 75\arcsec at 250, 350, and 500\,\micron\
respectively.  The camera consists of three silicon-nitride ``spider
web'' bolometer arrays \citep{turner2001} almost identical to those
for SPIRE on {\it Herschel} \citep{griffin2003}, with 149, 88, and 43
detectors at 250, 350, and 500\,\micron, organized in a hexagonal
close-packed pattern.  Radiation is coupled to the bolometers using
2$f\lambda$ spaced conical feed-horns, so that the $14' \times 7'$
field-of-view (simultaneously imaged by all three arrays) is
instantaneously under-sampled. The telescope must therefore scan in
order to produce fully-sampled images.  Scanning is also required to
modulate the bolometer signals to remove low-frequency detector drift.
A full description of the BLAST telescope and detectors is given in
\citet{pascale2007}. The 2005 Sweden flight performance is described
in \citet{truch2007}.

The optimal detector noise was obtained by scanning the telescope in
azimuth at 0.1\,deg\,s$^{-1}$ while drifting slowly in elevation such
that the scan lines are spaced 65\arcsec\ apart. In addition,
observations of Vulpecula were split between times when it was rising
and setting, resulting in cross-linked scans at angles of $\sim
45^\circ$. This cross-linking greatly reduces large-scale $1/f$ noise
in the map.  Details of the BLAST scanning technique are given in
\citet[][]{pascale2007}.

\subsection{Data Reduction} \label{sec:obsandred}

The raw BLAST05 data are reduced using a common pipeline detailed in
\citet{pascale2007} and \citet{truch2007}.  First, the 100\,Hz sampled
bolometer and gondola pointing data are de-spiked and the digital
filter responses are deconvolved.  Time-varying bolometer
responsivities are tracked using an internal calibration lamp. The
absolute gain of the instrument (including antenna efficiency),
determined from regular observations of Arp 220, is measured with
absolute uncertainties of 8\%, 10\%, and 12\% at 500, 350, and
250\,\micron\ respectively.  The relative pointing of the telescope
within a single map is determined to $<5''$ rms by integrating rate
gyroscopes with star trackers providing an absolute reference.
Residual pointing offsets between different observations of Vulpecula
are removed by direct rebinning of the data into maps, and aligning
the peaks of the seven brightest compact sources from each pass.
Finally, maps are made using a new algorithm called Signal And Noise
Estimation Procedure Including Correlations (SANEPIC). This technique
has evolved from strategies used to produce maps of the Cosmic
Microwave Background. A model is first developed for the raw bolometer
time-stream data in which the astronomical signal is produced by
projecting the estimate of the map into the time domain using the
known positions of each detector over time, and assuming that the
noise is Gaussian, stationary, and correlated in time and between
detectors.  The brightness of each map pixel is considered a free
parameter in this model. A maximum likelihood solution for the map is
then found using a direct inversion technique. This calculation
includes the full noise power spectra of the bolometers, and
propagates knowledge of correlations in the bolometer data to pixels
in the map.  Low-frequency noise (predominantly slowly-varying sky
emission) is naturally removed by this process.  In particular, to
reconstruct large spatial scales in the map, this method gives more
weight to data taken during a single visit, during which the bolometer
zero-point drift is minimal.  The DC level of the map is
unconstrained, and set to 0 by applying a weak high-pass filter to the
bolometer data before map-making.  The details of SANEPIC are given in
\citet{patanchon2007}.

\subsection{Overview of the Maps} \label{sec:mapoverview}
\begin{figure} 
\centering
\includegraphics[width=0.95\linewidth]{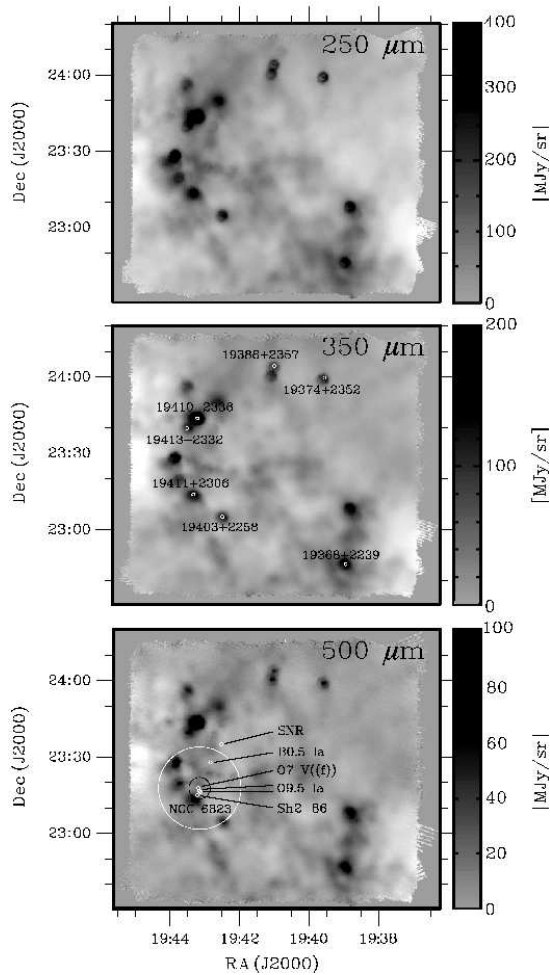}
\caption{Raw BLAST05 maps of Vulpecula produced with SANEPIC
  \citep{patanchon2007}. The similarity between the bands is due to
  the filters sampling primarily the Rayleigh-Jeans spectrum of
  thermal dust emission.  Previously identified {\it IRAS} sources
  associated with high-mass star forming regions are indicated on the
  350\,$\mu$m map (see \S\ref{sec:mapoverview}). It is assumed
  that most of the submillimeter emission originates in a molecular
  cloud complex at the distance of the open cluster NGC~6823
  ($\sim2.3$\,kpc away) indicated in the 500\,\micron\ map (the two circles
  give the cluster core size, and maximum extents, respectively).  The
  locations of several OB supergiants, an O7 V ((f)) star, and a
  supernova remnant, all associated with NGC~6823, are also shown.
  Sh2-86 is an extended \ion{H}{2} region excited by these massive
  stars.}
\label{fig:blastmaps}
\end{figure}
The BLAST05 maps of Vulpecula are shown in Figure~\ref{fig:blastmaps}.
The similarity of the three maps is striking, both in the diffuse
emission and the peaks, demonstrating that the bulk of the
measurements fall on the Rayleigh-Jeans tail of the thermal emission.
In this case, they are a good measure of dust column density in the
molecular cloud and the mass of the compact sources.
\begin{deluxetable}{ccrrr}
\tablewidth{0pt}
\tablecaption{Vulpecula map sensitivities.  \label{tab:mapsens}}
\tablehead{
\colhead{Band} &
\colhead{FWHP} &
\colhead{$\sigma_\mathrm{b}$} &
\colhead{$\sigma_\mathrm{p}$} &
\colhead{$\sigma_\mathrm{s}$} \\
\colhead{(\micron)}  &
\colhead{(\arcsec)} &
\colhead{(\arcsec)} &
\colhead{(mJy)} &
\colhead{(Jy)}
}
\startdata
\multicolumn{5}{c}{\it Raw maps} \\
250 & 207 & 88.2 &  18.0 & 0.32 \\
350 & 206 & 87.8 &   9.9 & 0.17 \\
500 & 212 & 90.4 &   9.9 & 0.18\vspace{1 ex}\\
\multicolumn{5}{c}{\it Deconvolved maps} \\
250 &  40 & 17.0 & 350.0 & 1.17 \\
350 &  50 & 21.3 &  99.0 & 0.41 \\
500 &  60 & 25.5 & 122.0 & 0.61 \\
\enddata
\tablecomments{Full-Width Half-Power (FWHP) is the diameter in which
  50\% of the beam power is contained (equivalent to Full-Width
  Half-Maximum for Gaussian beams).  The standard deviation (size) of
  a Gaussian beam with the equivalent FWHP is given by
  $\sigma_\mathrm{b}$. The rms noise in a single 18\arcsec\ map pixel
  is given by $\sigma_\mathrm{p}$. The rms noise in the equivalent
  FWHP Gaussian beam is given by $\sigma_\mathrm{s}$.  Values are
  presented both for ``Raw maps'' produced with SANEPIC
  (\S\ref{sec:mapoverview}) and for ``Deconvolved maps''
  (\S\ref{sec:decon}).}
\end{deluxetable}

The far-infrared (60 and 100\,\micron) emission in this part of
Vulpecula is dominated by several luminous high-mass star forming
regions.  Three of these have been studied recently by \cite{bel06}
and \citet{zha05}, associated with the following IRAS sources (for
later cross-reference we append the BLAST name from
Table~\ref{tab:iras}): 19368+2239 (V03); 19374+2352 (V05); and
19388+2357 (V08).  A further four regions have been studied
extensively by \cite{beu02}, using CS multi-line, multi-isotopolog
observations and the 1.2\,mm dust continuum: 19403+2258 (V18);
19410+2336 (V30); 19411+2306 (V32); and 19413+2332 (V40).
Note that these do not exhaust the list of bright sources, either
at submillimeter or far-infrared wavelengths.

In the submillimeter, the brightest high-mass star forming region near
IRAS~19410+2336 is visible in the north-east part of the BLAST maps
(Figure~\ref{fig:blastmaps}). The other bright regions are located on
a roughly ``C''-shaped arc readily apparent in the {\it IRAS} maps.
As with {\it IRAS}, the BLAST emission is in most cases centrally
peaked, representing a single massive clump at the observed resolution
of BLAST.  There is also diffuse emission from extended clouds of
dust, as well as blending of (clustered) point sources convolved with
the beam.

In the 2005 flight, the point spread function (PSF) was not
diffraction-limited (40\arcsec, 58\arcsec, and 75\arcsec\ at 250, 350,
and 500\,\micron\ respectively) as originally designed. The resulting
full-width half-powers (FWHP, equivalent to the full-width
half-maximum for a Gaussian) in all three bands are approximately
$3\farcm5$ and exhibit complex shapes that vary as a function of
wavelength and position in the focal plane \citep{truch2007}.

The beam shape is clearly recognizable at many locations across the
entire field, suggesting the presence of point-like objects with
angular scales smaller than the beam. However, the overlapping beam
patterns also indicate that the maps are highly source-confused
(Figure~\ref{fig:blastmaps}).  Even though only 3--8\% of the beam
power is found in the central diffraction-limited peak at each
wavelength, the signal-to-noise (\snr) of the maps are high enough
(compare sensitivity for ``Raw maps'' in Table~\ref{tab:mapsens} with
typical source flux densities in later tables) that we are able to
deconvolve a significant fraction of the beam to recover angular
scales close to the diffraction limit.  We can then use these
deconvolved maps to search for compact sources.

\subsection{Image Deconvolution}
\label{sec:decon}

Deconvolution in astronomy has a long history \citep[see the review
by][]{starck2002}. The convolution problem can be stated, using the
notation of \citet{starck2002}, as

\begin{equation}
\label{eqn:convprob}
I(x,y) = (O \ast P)(x,y) + N(x,y),
\end{equation}

\noindent where $I$ is the observed map, $O$ is the true image, $P$ is
the instrument's PSF, $N$ is measurement noise, and ``$\ast$'' is the
convolution operator. Invoking the convolution theorem, this
expression can be written in Fourier space as

\begin{equation}
\hat{I}(u,v) = \hat{O}(u,v) \hat{P}(u,v) + \hat{N}(u,v) ,
\end{equation}

\noindent where $\hat{X}$ is the Fourier transform of $X$. A naive
solution to the convolution problem is found by simple division in
Fourier space,

\begin{equation}
\label{eqn:convsol}
\hat{\tilde{O}}(u,v) = \frac{\hat{I}(u,v)}{\hat{P}(u,v)} = 
  \hat{O}(u,v) + \frac{\hat{N}(u,v)}{\hat{P}(u,v)}.
\end{equation}

However, in practice, this direct inversion method amplifies the noise
at high spatial frequencies. Many methods have been developed to solve
Equation~\ref{eqn:convprob} iteratively, including least squares,
maximum likelihood and wavelet-based methods.  These algorithms
typically require high-precision knowledge of the PSF, which we do not
have for the BLAST05 maps. The effective PSFs vary significantly
across the maps due to asymmetries in the beam-pattern (which itself
is well-measured).  Over time the orientation of this pattern on the
sky changes, such that the effective shape of a point source depends
on the amount of time spent observing at different parallactic angles,
and instrumental noise variations. Iterative methods for deconvolution
were attempted, but the variable PSFs proved to be too problematic. We
concluded that strong artifacts due to convolution are inevitable and
we decided to use the direct inversion method, even though it is
probably non-optimal.

Equation~\ref{eqn:convsol} can be rewritten in real space as
$\tilde{O} = I \ast K$, where $K$, the deconvolution kernel, is the
inverse transform of $\hat{P}^{-1}$. In order to suppress the
amplification of noise at high frequencies we effectively re-convolve
the map by a Gaussian, $G$, with width approximately equal to the
designed diffraction limit of the telescope. This can be combined with
the deconvolution formulation by writing $\hat{K} = \hat{G} /
\hat{P}$. Here $\hat{K}$ is also smoothed at large spatial frequencies
to further suppress noise spikes as $\hat{P}$ goes to zero.
Additionally, spikes in $\hat{K}$ due to zero-crossings in $\hat{P}$
are clipped.

\begin{figure} 
\centering
\includegraphics[width=\linewidth]{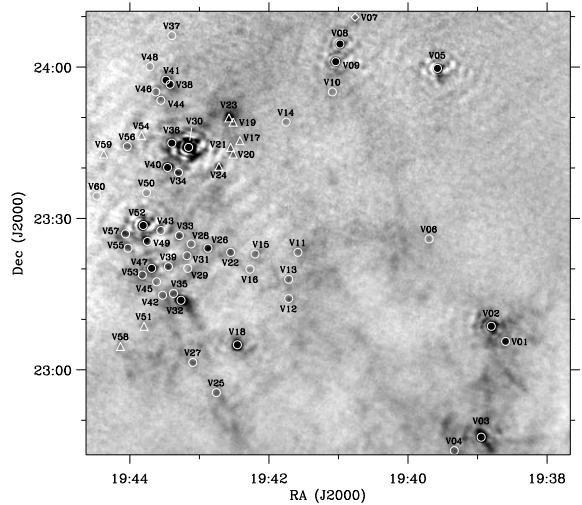}
\caption{The grayscale image is a resolution enhanced BLAST
  350\,\micron\ map, with the locations of compact sources
  (Table~\ref{tab:src}) indicated: 49 objects associated with the
  clouds surrounding NGC~6823 at $\sim$2.3\,kpc (circles); 10 objects
  in the Perseus Arm at $\sim$8.5\,kpc (triangles); and a single
  outer-galaxy object at $\sim$14\,kpc (diamond) --- see
  \S\ref{sec:dist} for a discussion of distance estimates.  The
  resolution enhancement was accomplished with direct Fourier
  deconvolution of the PSF (see \S\ref{sec:decon}).  Residual
  ringing is an artifact of uncertainties in the beam shape which
  varies slightly across the map.}
\label{fig:decon}
\end{figure}

The PSFs are estimated by two methods. In the first case, a synthetic
PSF is constructed based on measurements of point sources throughout
the flight. The synthetic PSF attempts to account for the range of
parallactic angles over which the field was observed by averaging
together appropriately-weighted beam rotations. The second method uses
a point source directly from the map. An isolated source from the
north-west corner of the map is used (V05 in Figure~\ref{fig:decon}),
this having relatively low surrounding diffuse emission. A low-order
polynomial is fit to the background and subtracted from the source to
remove the diffuse emission.  The deconvolutions are performed using
both types of PSF.  The synthetic PSFs provide the best results at 350
and 500\,\micron, while the isolated source is better for the
250\,\micron\ map, based on the amplitude of the ripples seen in the
resulting deconvolved maps.

In order to reduce edge effects caused by the Fourier transforms, both
the maps, $I$, and the PSFs, $P$, are apodized and zero-padded prior
to deconvolution. The map is apodized over a scale of 5\arcmin\ along
a rectangle bordering the map and the PSF is apodized over 2\arcmin\
along a circle of radius 4\arcmin.

Finally, the variance map produced by SANEPIC is propagated through
the deconvolution filter, providing a noise map for use in
source-finding and fitting. The noise at each wavelength both before
and after deconvolution are given in Table~\ref{tab:mapsens} under
``Raw maps'' and ``Deconvolved maps'' respectively.

\subsection{Compact Source Identification}
\label{sec:sourceid}

We search for individual compact sources in the deconvolved maps using
source-finding analysis based on the use of a ``compensated PSF'',
also called the Mexican Hat Wavelet (MHW) technique \citep[see
e.g.,][]{barnard2004}, which identifies objects in confused images by
subtracting a local background. We apply the MHW with a characteristic
width equal to the nominal resolution (the width of $G$ in the
deconvolution), to the BLAST deconvolved maps. The peaks of all $\sim
6$-$\sigma$ clumps in each waveband are identified as potential
sources. Due to the excess noise produced by the deconvolution
process, many false detections are found near bright sources.  No
robust method to automatically reject these false peaks was found,
since the ripples are not uniform across the image. We visually reject
any peak that appears to be associated purely with the noise ripples.
In addition, 11 of the recovered sources that lie slightly further
away from the brightest objects remain clearly affected by these
residual ripples. Since the ripples are different in each band the net
effect is to contaminate the observed BLAST colors.  In total,
$\sim50$\% of the rejected sources have formal statistical
significances below $9$-$\sigma$, although there is a tail of rejected
sources extending to $\sim25$-$\sigma$ that have relatively higher
systematic errors since they lie on the largest ripples next to the
brightest submillimeter objects.  The positions of BLAST sources
derived from the deconvolved maps are shown in Figure~\ref{fig:decon}.

The list of robust detections in each waveband are combined and a
circular region around each peak in the three maps is then
simultaneously fit with a single Gaussian. In addition nearby sources
are fit using multiple Gaussians. In each region, a fourth order
polynomial baseline is fit to the surrounding regions and subtracted.
The amplitudes in each waveband, position, and width are all free
parameters in the fit. The fit is calculated using a non-linear
least-squares minimization routine. Constraints on source position and
width are included to reduce the possibility of divergence in the
fitting procedure. Leaving the source width as a free parameter in the
fit biases the resulting flux densities high, but is necessary since
the beam varies significantly across the field.  This bias is
investigated in the next section.  Given the \snr\ of these sources,
however, even after deconvolution, the flux density uncertainty is
dominated by the calibration error.  The catalog of source positions
and BLAST flux densities are given in Table~\ref{tab:src}.

\subsection{Monte Carlo Simulations}
\label{sec:montecarlo}

In order to estimate the effectiveness of our compact source
extraction process, a series of Monte Carlo simulations are performed.
Flux densities, biases, completeness estimates, and positional error
distributions are investigated by inserting a point source, convolved
with the estimate of the beam, into the actual BLAST maps produced
with SANEPIC. The deconvolution and source extraction processes are
applied and the extracted flux densities compared with the inputs. The
source extraction routine is modified slightly from that described in
\S\ref{sec:sourceid} in order to avoid manually rejecting unreliable
sources in residual ripples around bright sources --- a simple
automatic routine was implemented that is able to reproduce the manual
rejection procedure for the original source list to an accuracy of
$\sim 10$\%.

\begin{figure}
\centering
\includegraphics[width=.8\linewidth]{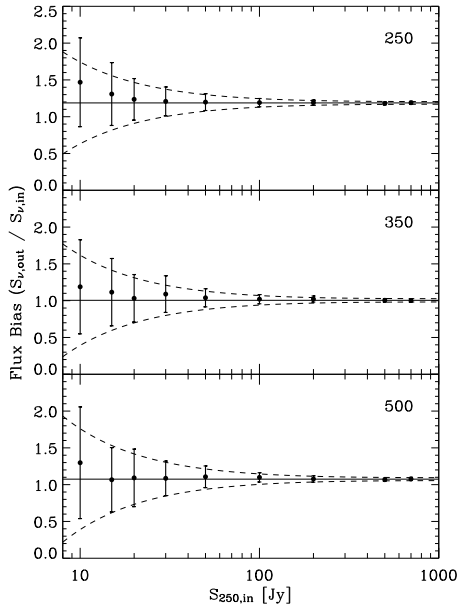}
\caption{The flux density bias, determined from 500 Monte Carlo
  simulations, is shown for each waveband. The bias is modeled as a
  constant function of input source flux density and the errors are
  modeled with two components (described in
  \S\ref{sec:montecarlo}).}
\label{fig:blastfluxsims}
\end{figure}

\begin{figure}
  \centering
  \includegraphics[angle=90,width=\linewidth]{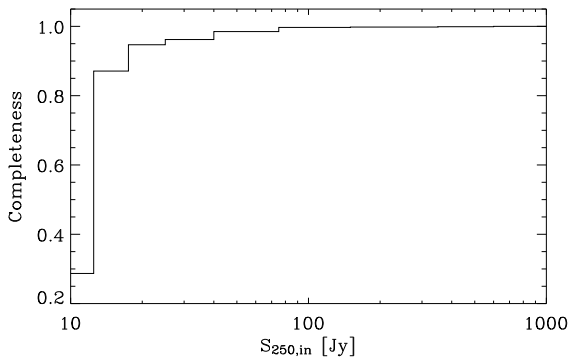}
  \caption{The completeness as a function of 250\,\micron\ flux
    density determined from 1000 Monte Carlo simulations (described in
    \S\ref{sec:montecarlo}).}  
  \label{fig:blastcomplsims}
\end{figure}

In order to extract the flux density bias and errors, 500 iterations
are performed at each of several input flux densities spanning the
range of source brightnesses found in the map.  At each iteration, a
250\,\micron\ flux density is chosen and the 350 and 500\,\micron\
flux densities are assigned based on the median source colors found in
the field. A simulated source is considered detected if an additional
peak is found compared to the number of sources originally found in
the map. In some cases, the simulated source will by chance land on
top of a real source at the same position. In such cases, the
detection of the simulated source is only counted if the flux density
exceeds that of the real source that it obscures. The resulting output
flux density distributions are roughly Gaussian with a positive tail.
The Bayesian 68\% confidence limits are found and the distributions
are fit by the Gaussian that passes through these points. We find that
the measured flux density is biased high by a factor of 1--1.2,
independent of input flux density (Figure~\ref{fig:blastfluxsims}).
Based on these results, we can describe the spread in flux density
with a 2-component model: the first is independent of flux density,
reflecting noise in the map, and the other is linear in flux density,
due to uncertainty in the fitting process.  Positional errors are
calculated using the same simulations. These errors are described by
circular Gaussians with $\sigma$ ranging from $\sim 7$\arcsec\ at the
bright end to $\sim 70$\arcsec\ at the faint end.  
Completeness is estimated using the same set of simulations, except
that the flux density fits are not performed. At each of the input
flux densities, 1000 simulations are performed and the number of times
that the input is detected is counted. The results are shown in
Figure~\ref{fig:blastcomplsims}.

In total, 60 compact sources are detected in the Vulpecula maps. Flux
densities and their uncertainties are given in Table~\ref{tab:src}.
Note that the tabulated values include a color correction based on SED
fits; see \S\ref{sec:colorcorrect} for a description of this
procedure.  Figure~\ref{fig:decon} indicates the positions of the
detected sources.
\section{INFRA-RED PHOTOMETRY}
\label{sec:firsed} 

The BLAST data alone cannot accurately constrain the SED corresponding
to thermal emission in many of the sources from this survey.  At
temperatures $T\ga25$\,K (and $\beta=1.5$) all of the BLAST filters
sample the Rayleigh-Jeans tail. To measure the luminosity and dust
mass, additional photometry at slightly shorter wavelengths
($\sim$50--100\,\micron) is required to identify the emission peak.
FIR measurements and upper limits are estimated for all of the sources
using {\it IRAS} 60 and 100\,\micron\ photometry and the new MIPS
70\,\micron\ map from MIPSGAL \citep{car05}.  Although most of the
luminosity is produced at submillimeter and FIR wavelengths, a
non-negligible portion is also observed in the mid-infrared (MIR)
($\lambda\sim$10--50\,\micron). This radiation is emitted from hotter
dust that lies close to the heat source.  Protostars produce
significant amounts of radiation in this band, since the material in
the vicinity of the object is optically thick. This material is
irradiated with UV light which is then re-emitted at longer
wavelengths.  To probe the MIR SEDs of BLAST sources we used {\it
  IRAS}\/ 12 and 25\,\micron\ measurements and the {\it MSX}
\footnote{\url{http://www.ipac.caltech.edu/ipac/msx/msx.html}} catalog
at 8, 12, 14, and 21\,\micron.

\subsection{{\it IRAS} 12, 25, 60, and 100\,\micron\ Flux Densities}
\label{sec:iras}

A significant fraction of the BLAST sources have clear counterparts in
the {\it IRAS} Point Source Catalog version 2.0
\citep[PSC,][]{helou1988}. We add the BLAST positional uncertainties
in quadrature with the semi-major axes of the PSC error ellipse to
determine search radii for each source. The BLAST uncertainties are
taken from the simulations described in \S\ref{sec:montecarlo},
but we conservatively set the error at the bright end to 30\arcsec.
Identifications from the {\it IRAS} PSC were found for 23 of the 60
BLAST sources, as summarized in Table~\ref{tab:iras}.

For sources that lack PSC counterparts or measurements in any of the
{\it IRAS} bands, we produce measurements or upper limits directly
from {\it IRAS} maps. We use the {\it IRAS} Galaxy Atlas
\citep[IGA,][at 60 and 100\,\micron]{cao1997} and the Mid-Infrared
Galaxy Atlas \citep[MIGA,][at 12 and 25\,\micron]{kerton2000}. These
maps are produced using the resolution-enhancing algorithm HIRES
\citep{aumann1990}. The PSFs vary across these maps, showing strong
elongation along the scan direction. However, these HIRES maps resolve
sources that are also detected with BLAST (see
Figure~\ref{fig:irascompare}) yet appear confused in {\it IRAS} maps
that have not undergone resolution enhancement \citep[e.g., the IRIS
maps of][]{miv2005}.

We use aperture photometry to measure flux densities in the IGA and
MIGA maps. A circular aperture radius of 2\farcm4 is used at
100\,\micron.  For the remaining bands an elliptical aperture aligned
with the scan direction is used, with semi-axes $1\farcm8 \times
1\farcm2$. These apertures are centred over the BLAST coordinates and
the maps are integrated.  Baseline pixel values are estimated as the
median in an annulus between the outer-edge of the photometry aperture
and a second circle (or ellipse) that is larger by a factor of 1.3.
This size was chosen through trial-and-error as a compromise between
smaller sizes which encompass fewer pixels (baseline estimates with
larger statistical errors), and larger sizes which suffer greater
contamination from large-scale extended structure or adjacent sources
(baseline estimates with larger systematic errors).  Since many of the
sources remain confused, the flux density and baseline measurements
are both checked visually. In cases where a small amount of emission
is detected in the baseline annulus, the confused fraction is excised
in the estimates.  Furthermore, if additional sources are seen near
the edge of the measurement aperture, or if it appears to lie on a
bright gradient of background emission, the measurement is flagged as
an upper-limit.  Finally, if no source is visible in the aperture, it
is flagged as a non-detection. The uncertainty in all of the
measurements is estimated as the rms in the non-detections. The
results of this photometry procedure are summarized in
Table~\ref{tab:mapphot}.

To verify our technique, we compare these aperture measurements with
flux densities from the {\it IRAS} PSC and find no significant bias,
with scatters of 12\% at 60\,\micron\ and 25\% at 100\,\micron.

\subsection{MIPS 70\,\micron\ Flux Densities}
\label{sec:mips70}

After examining the {\it IRAS} PSC, and measuring flux densities in the
IGA maps, 32 of the 60 BLAST sources lacked a clear detection at
either 60 or 100\,\micron. To constrain the FIR SEDs in these cases,
we use mosaics of the 70\,\micron\ MIPSGAL \citep{car05} images
downloaded from the {\it Spitzer} public web site using the {\it
  Spitzer} Pride
software.\footnote{\url{http://ssc.spitzer.caltech.edu/propkit/spot/}}
The 8\,$\mu$m IRAC image (Figure~\ref{fig:iracblast}) is also produced
using this software.

Similar to the flux densities obtained from {\it IRAS} maps in
\S\ref{sec:iras}, aperture photometry with baseline correction is used
for the 70\,\micron\ map. The radius of the measurement aperture is
1\arcmin\ and the baseline aperture is an annulus with inner and outer
radii of 1\arcmin\ and 1\farcm3. It was noticed that these
measurements yield flux densities systematically lower than the
60\,\micron\ observations, despite the fact that the SED for every
source should be brighter at 70\,\micron, since these wavelengths fall
on the Wien tail of the thermal emission in all cases. This problem
arises because saturation and non-linearities affect MIPS 70\,\micron\
sources with flux densities $\ga 20$\,Jy (A.  Noriega-Crespo, private
communication).  Fortunately only the fainter BLAST sources require
70\,\micron\ photometry to constrain their FIR SEDs, and adopting a
20\,Jy cut in the catalog of 70\,\micron\ measurements yields useful
data for 10 BLAST sources with no {\it IRAS} detections. The
uncertainty in this photometry was estimated as 3\,Jy from the rms of
observations that were visually flagged as non-detections.  A
comparison between MIPS, {\it IRAS}, and BLAST is shown in
Figure~\ref{fig:irascompare}, and the MIPS photometry is summarized in
Table~\ref{tab:mapphot}.

\subsection{500--8\,\micron\ Map Comparison}
\label{sec:mapcompare}

\begin{figure}
\centering
\includegraphics[width=\linewidth]{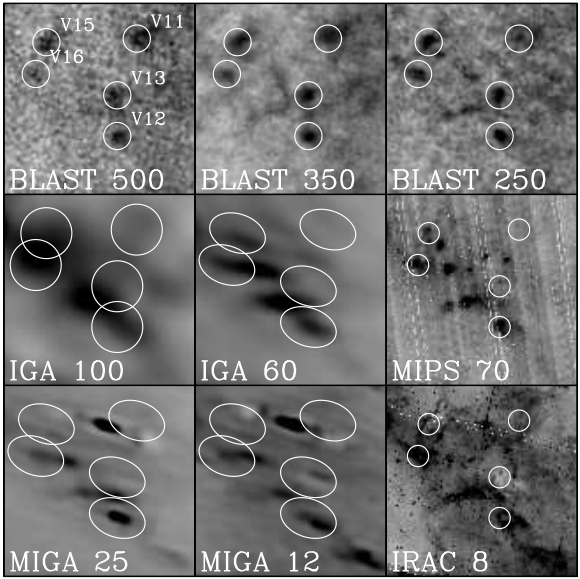}
\caption{Comparison between BLAST05 deconvolved maps, {\it IRAS} maps
  produced with HIRES, and the {\it Spitzer} MIPS (70\,\micron) and
  IRAC (8\,\micron) maps. Each thumbnail has a size of $0\fdg3 \times
  0\fdg3$. Circles on the BLAST maps and IRAC map indicate source
  positions with arbitrary diameters 2\farcm5 and 2\arcmin\
  respectively.  The circles and ellipses indicated on the {\it IRAS}
  and MIPS maps are the apertures that were used for photometry. The
  SED of the coldest object, V11, is shown in
  Figure~\ref{fig:mergedSED2}.}

\label{fig:irascompare}
\end{figure}

Figure~\ref{fig:irascompare} shows an expanded view of a $0\fdg3
\times 0\fdg3$ region of the maps described in the previous sections,
encompassing the BLAST sources V11, V12, V13, V15, and V16.  It is
clear that the relative brightnesses of sources vary greatly over the
range 500--8\,\micron, due to their different intrinsic SEDs.  None of
the BLAST sources in this region of the map have counterparts in the
{\it IRAS} PSC\@.  Source confusion is also an issue; for example,
although V12 is not in the {\it IRAS} PSC, a counterpart is seen in
the higher resolution IGA 60\,\micron\ and the MIPS 70\,\micron\
images (and MIGA as well).  V16 has a faint counterpart at
70\,\micron.  In the same images, {\it IRAS} PSC sources not seen by
BLAST appear: 19395+2313 between V11 and V13, 19397+2309 to the NE of
V12, and 19399+2312 W of V16; the source SW of V15 is not in the {\it
  IRAS} PSC.  Based on the brightness and color of 19397+2309 at 60
and 100\,\micron\, (23 and 80\,Jy respectively), we would expect to
have detected this source with BLAST.  Comparison of the 250\,\micron\
image with the 70\,\micron\ image (where the source is resolved into
two), suggests a detection by BLAST, but it did not meet our
compactness criteria to be tabulated as a point source.  At even
shorter wavelengths, additional sources and nebulosities appear, for
example 19397+2315 E of V11.

\subsection{{\it MSX} 8, 12, 14, and 21\,\micron\ Flux Densities}
\label{sec:msx}

{\it MSX} flux densities for the BLAST sources are obtained from a
cross-correlation with the {\it MSX} Point Source Catalog version 2.3
\citep{egan2003}.  As with the {\it IRAS} PSC
(\S\ref{sec:iras}), {\it MSX} counterparts are identified within
a search radius of the BLAST positions that varies as a function of
the source brightness. The positional uncertainties in the {\it MSX}
catalog are negligible by comparison (rms $\sim4$--5\arcsec).  We find
potential {\it MSX} counterparts for 40 objects in the BLAST catalog.
Inspection of the {\it MSX} maps shows that some of these sources are
found in regions with diffuse emission, and, given the angular
resolution of the BLAST catalog, we cannot presently determine whether
all of the proposed {\it MSX} associations are BLAST objects, or other
protostars in the same star forming region at an older evolutionary
stage. In cases where a single {\it MSX} counterpart is identified we
assume that it is in fact the same source. In cases with multiple
proposed {\it MSX} counterparts, however, the sum of the flux
densities is taken as an upper limit. Presently no attempt has been
made to use the shape of the {\it MSX} SEDs to justify their
associations with the BLAST objects. All candidates are given in
Table~\ref{tab:msx}.

\section{SUBMILLIMETER--MIR SEDS} 
\label{sec:coldsed}

With submillimeter--MIR photometry, we fit the temperature and
bolometric flux produced by the coldest dust, as well as the
bolometric flux produced in the MIR.  We note that all of our sources
are likely composed of regions at different temperatures, typically a
warmer core embedded in a colder and less dense medium.  Our goal is
to use a simple SED model as an interpolation function for the
sparsely sampled photometry to estimate the total luminosity of each
source, and to infer an approximate temperature for the dominant
emission from cold dust.

We assume optically-thin emission from an isothermal modified
blackbody,
\begin{equation}
  S_{\nu} = A
  \left(\frac{\nu}{\nu_0}\right)^{\beta} B_{\nu}(T),
\label{eq:sed}
\end{equation}
where $A$ is the amplitude of the SED, $B_{\nu}(T)$ is the Planck
function, $\beta$ is the dust emissivity index, and the emissivity
factor (brackets) is normalized at a fixed frequency $\nu_0$.  Note
that the amplitude can be expressed in terms of a total clump mass,
$M_{\rm c}$, the dust mass absorption coefficient $\kappa_0$
(evaluated at $\nu_0$), and the distance to the object, $d$,
\begin{equation}
A = \frac{M_{\rm c} \kappa_0}{Rd^2}.
\label{eq:mass}
\end{equation}
Since $\kappa_0$ refers to a dust mass, the gas-to-dust mass ratio,
$R$, is required in the denominator to infer total masses.  We adopt
$\kappa_{0} = 10$\,cm$^2$\,g$^{-1}$, evaluated at
$\nu_0=c/250$\,\micron, from \citet{hildebrand1983}. A factor of 100
is assumed for $R$. We note that the combined mass uncertainty due to
$\kappa_{0}$ and $R$ is at least as large as a factor $\sim2$
\citep[e.g.,][]{hildebrand1983,ossenkopf1994,kerton2001}, depending on
assumptions about the environment, age, chemical composition, and
shape of the dust grains.  Note that our adopted value, along with
$\beta = 1.5$, is equivalent to the 1.2\,mm opacity adopted by
\citet{motte2007}.

Equation~\ref{eq:sed} is fit to all of the available photometry from
500--60\,\micron\ using $\chi^2$ optimization (except for some cases
where the FIR peak is very broad, in which case it is only fit from
500--100\,\micron). As in \citet{truch2007} the band-averaged flux
density of the model SED is calculated with knowledge of the BLAST
filter passbands before comparing them with measurements. We also
include the correlated calibration uncertainties with the statistical
uncertainties estimated in \S\ref{sec:montecarlo}.  The data
covariance matrix, $C$, is constructed by placing the estimated
variance for each data point along the diagonal. For the BLAST data
points, the diagonals are calculated as the quadrature sum of the
statistical uncertainties with the calibration uncertainties listed in
Table~1 in \citet{truch2007}. The off-diagonal cross-correlation terms
are estimated directly from the Pearson correlation coefficients in
the same table.  $\beta$ is not well constrained and is therefore
fixed to $\beta=1.5$ (consistent with Figure~\ref{fig:blastcolcol}) so
that only $A$ and $T$ are allowed to vary. For the 11 cases where the
BLAST colors are unreliable (\S\ref{sec:sourceid}; marked in
Table~\ref{tab:src}), a temperature of 20\,K has been adopted.

\subsection{Including Upper Limits in $\chi^2$}
\label{sec:upperlim}

Many of the fainter sources have only BLAST detections. In order to
make the best use of FIR upper-limits, ``survival analysis'' is used
to include the upper-limits in the calculation of $\chi^2$.  Given a
measurement that is deemed an upper limit, with flux density,
$s_{\mathrm{l}}$, and uncertainty, $\sigma_{\mathrm{l}}$, the likelihood
of the model flux density, $\tilde{s}$, is obtained from the integral of
the appropriate tail of the likelihood of a detection,
\begin{equation}
\label{eqn:survival}
  p(\tilde{s}|s_{\mathrm{l}},\sigma_{\mathrm{l}}) = 
  \frac{1}{\sqrt{2\pi}}\int^\infty_{(\tilde{s}-s_{\mathrm{l}})/\sigma_{\mathrm{l}}} e^{-z^2/2} dz,
\end{equation}
where we have assumed Gaussian noise.  Equation~\ref{eqn:survival} is
the ``survival function'' for censored data \citep{isobe1986}. The
behavior of this function is easily seen with a few examples.  If the
model flux density is much smaller than the measured limit, then the
likelihood goes to 1. If the model is equal to the measured limit, the
likelihood is 0.5. The likelihood of models more than a few $\sigma$
above the limit drops to $\sim0$.

Noting that $\chi^2$ is the negative log-likelihood function, and
assuming the likelihood of the upper limits is independent of the
detections, then
\begin{equation}
\label{eqn:chisq}
  \chi^2 = (\tilde{s}-s_{\mathrm{d}})C^{-1}(\tilde{s}-s_{\mathrm{d}})^T - 
  \ln p(\tilde{s}|s_{\mathrm{l}},\sigma_{\mathrm{l}}).
\end{equation}
Here $s_{\mathrm{d}}$ are the measured detections associated with the
data covariance matrix, $C$. Equation~\ref{eqn:chisq} gives the
expression that we minimize in order to fit the region of the data
consistent with the isothermal SED model ($\ga 60$\,\micron).

\subsection{SED Fits}
\label{sec:sedfits}

Uncertainties for the model parameters $A$ and $T$, and non-linearly
dependent quantities such as the FIR integrated flux, $M_\mathrm{c}$,
and their distance-independent ratio $L_{\mathrm{FIR}}/M_\mathrm{c}$,
are obtained from Monte Carlo simulations.  Mock data sets are
generated from realizations of Gaussian noise, including both
correlated and uncorrelated errors, as described by the covariance
matrix. The $\chi^2$ minimization process is repeated for each data
set, and the resulting parameters and dependent quantities placed in
histograms.  Means and 68\% Bayesian confidence intervals measured
from the relevant histograms are given in Table~\ref{tab:sed}.  Note
that errors in $T$ have opposite effects on $L_{\mathrm{FIR}}$ and
$M_\mathrm{c}$, which exaggerates the uncertainty in
$L_{\mathrm{FIR}}/M_\mathrm{c}$.

\begin{figure}
\includegraphics[width=\linewidth]{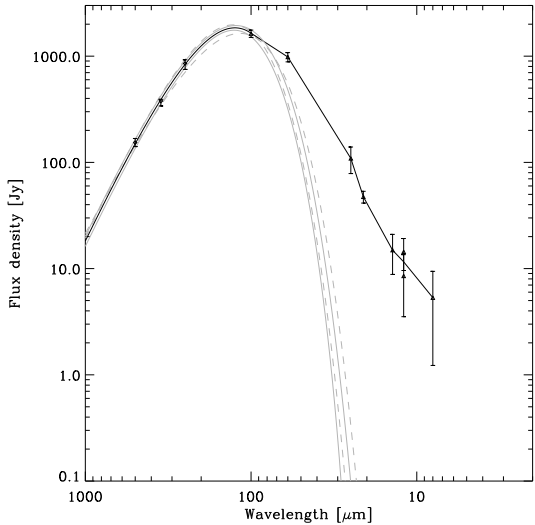}
\caption{Example SED for V30, the brightest BLAST object
  ($T=25.9$\,K).  BLAST error bars include correlated calibration
  uncertainties ($\sim$10\%). Gray lines show the 68\% confidence
  envelope of modified blackbody models from Monte Carlo simulations
  in which the dust emissivity index is fixed at $\beta=1.5$. Dashed
  gray lines are particular realizations from a second Monte Carlo
  simulation in which $\beta$ is also left as a free parameter, here
  indicating the 1-$\sigma$ envelope of $T$ (\S\ref{sec:sedfits}).
  The black line shows the best-fit modified blackbody connected
  continuously at 100\,$\mu$m to a series of piecewise-continuous
  power-laws between the shorter-wavelength data.  }
\label{fig:mergedSED}
\end{figure}

\begin{figure}
\includegraphics[width=\linewidth]{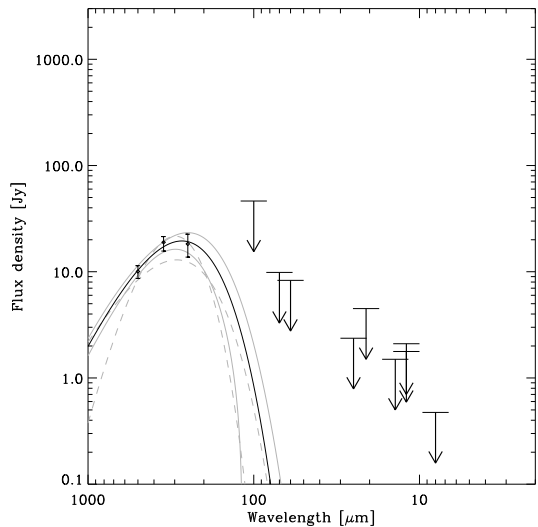}
\caption{Example SED for V11, the coldest BLAST source ($T=12.3$\,K).
  Symbols have the same meaning as in Figure~\ref{fig:mergedSED}.  The
  FIR spectrum clearly turns over in the BLAST bands, and there are no
  detections at the {\it IRAS}, MIPS or {\it MSX} wavelengths (see
  Figure~\ref{fig:irascompare}). The bolometric flux for this source
  can only be crudely estimated from the integral of the modified
  blackbody.}
\label{fig:mergedSED2}
\end{figure}

Figure~\ref{fig:mergedSED} shows an example SED for V30, the brightest
BLAST object in the sample. The solid gray lines indicate the 68\%
confidence envelope of modified blackbodies that fit the BLAST and
{\it IRAS} 100\,\micron\ data. The black line consists of the best-fit
modified blackbody at wavelengths $> 100$\,\micron, and a series of
piecewise-continuous power-laws through the {\it IRAS} and {\it MSX}
data points at wavelengths $\le 100$\,\micron. The best fit
temperature to the submillimeter data is 26\,K, considerably smaller
than 46\,K estimated from MIR and mm data away from the peak
\citep{Sridharan02}.  The luminosity fraction given in column 5 of
Table~\ref{tab:sed} is the ratio (as a percentage) of the FIR
integrated flux from the best-fit modified blackbody (solid gray
lines) to the total bolometric flux, estimated from the integral of
the black line across the wavelength range 2--5000\,\micron.
Similarly, Figure~\ref{fig:mergedSED2} shows the SED of V11, the
coldest object in the sample ($T=12.3$\,K). As seen in this example,
observations across the BLAST wavelength range clearly reveal a
turnover in the FIR SED for the coldest objects. The lack of
counterparts in the shorter-wavelength bands is emphasized in
Figure~\ref{fig:irascompare}.

The range of modified blackbody models from the Monte Carlo
simulations are generally very tight (e.g.~for V30 $\Delta T =
1.0$\,K, and for V11 $\Delta T = 1.6$\,K). The narrowness of this
range is largely due to the prior constraint that the dust emissivity
index is precisely $\beta=1.5$. There is a strong degeneracy between
the values of $\beta$ and $T$; for example, at larger values of
$\beta$ the steepness of the Rayleigh-Jeans spectrum increases,
requiring lower values of $T$ to compensate for that steepness, by
bringing the FIR peak to longer wavelengths and hence closer to the
shape of the data reported in the BLAST bands. For the two examples
discussed here, we re-fit the $\lambda \ge 100$\,\micron\ region of
the SED, letting $A$, $T$, and $\beta$ vary independently in 100 Monte
Carlo simulations.  For V30 the resulting range of temperatures and
dust emissivities are $T=27.6\pm2.8$\,K, and $\beta=1.4\pm0.3$
(compared with $T=25.9\pm1.0$\,K when $\beta=1.5$).  For V11 the
ranges are $T=9.4\pm3.4$ and $\beta=2.9\pm1.5$ (compared with
$T=12.3\pm1.6$ when $\beta=1.5$). Particular realizations from these
Monte Carlo simulations have been selected to illustrate the
approximate 1-$\sigma$ range of $T$ as dashed gray lines in
Figures~\ref{fig:mergedSED} and \ref{fig:mergedSED2}. Many of the
BLAST sources have poorly constrained values of $T$ and $\beta$ as in
the example of V11. We have therefore chosen to fix $\beta=1.5$ for
the remainder of the analysis in this paper, which reduces inferred
errors in $T$ by factors of $\sim2$--3. It also emphasizes the range
of SEDs uncovered by BLAST, since under this constraint $T$ uniquely
determines the wavelength of the FIR peak.

In addition to the effect of the prior on $\beta$, the treatment of
correlated uncertainties between the BLAST measurements lead to
inferred temperature ranges that are perhaps smaller than what one
might expect given the sizes of the error bars in
Figures~\ref{fig:mergedSED} and \ref{fig:mergedSED2}.  For example,
V30 has photometric errors that are only $\sim2$\% in all three bands
(Table~\ref{tab:src}). However, the error bars shown in these figures
include approximately $\sim10$\% calibration uncertainties that are
nearly 100\% correlated across the three bands.  This type of
uncertainty only affects the absolute {\em scale} of the SED, $A$, but
not the {\em shape} which is encoded in $T$ and $\beta$ (for which
only the independent 2\% photometric errors are relevant). For this
reason the temperature uncertainty in V30 (under the model
assumptions) is only $\Delta T = 1.0$\,K, demonstrating the superior
ability of BLAST to constrain the shape of the FIR SED of cold sources
when systematic uncertainties are accounted for appropriately.

\subsection{Color-corrected BLAST Flux Densities}
\label{sec:colorcorrect}

\begin{figure}
\centering
\includegraphics[width=\linewidth]{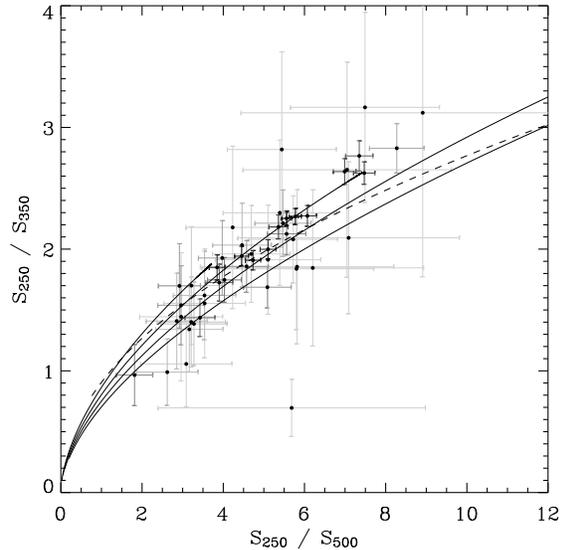}
\caption{Color-color plot for the compact sources detected by BLAST05.
  We have omitted several sources with unreliable colors, identified
  in Table~\ref{tab:src}, and two sources with highly unconstrained
  colors due to faint flux densities.  Modified blackbody models with
  constant $\beta$, equal to 0, 1, 2, and 3, increasing towards the
  bottom, and temperatures ranging from 5 to 200\,K are overplotted as
  solid lines. The dashed line represents the same model with a
  constant temperature of 20\,K and with $\beta$ ranging from $-1$ to
  3. The 1-$\sigma$ statistical error bars are shaded so that the more
  significant detections are darker. The flux densities plotted have
  been band-corrected as discussed in \S\ref{sec:colorcorrect}.}
\label{fig:blastcolcol} 
\end{figure}

BLAST filters have broad spectral widths that are $\sim 30$\% of the
central frequency \citep{pascale2007}. Since colors sampled by the
filters are a strong function of temperature for cooler objects
($T\la25$\,K), a color correction is required to enable direct
comparison with SEDs.  Once the SED (Equation~\ref{eq:sed}) has been
fit to data by minimizing $\chi^2$ (Equation~\ref{eqn:chisq}), we have
the choice of either calculating different effective wavelengths for
each measurement, or correcting each flux density at fixed
wavelengths. We choose the latter, and correct the flux densities to
precisely 250, 350, and 500\,\micron\ \citep[this procedure is
described in][]{truch2007}. These corrected flux densities are given
in Table~\ref{tab:src}, and a color-color plot for all 60 sources is
shown in Figure~\ref{fig:blastcolcol}. This plot illustrates the
variation in temperatures probed by the BLAST wavelengths, and
confirms that the choice of $\beta=1.5$ for the SED fits is close to
the center of the distribution. We have also tested the
color-correction method using $\beta=1$ and $\beta=2$ and find that
the effect on the measured flux densities is small compared with the
photometric uncertainties.

\subsection{Distance-independent Clump Properties}
\label{sec:dustproperties}

\begin{figure}[t]
\includegraphics[width=\linewidth]{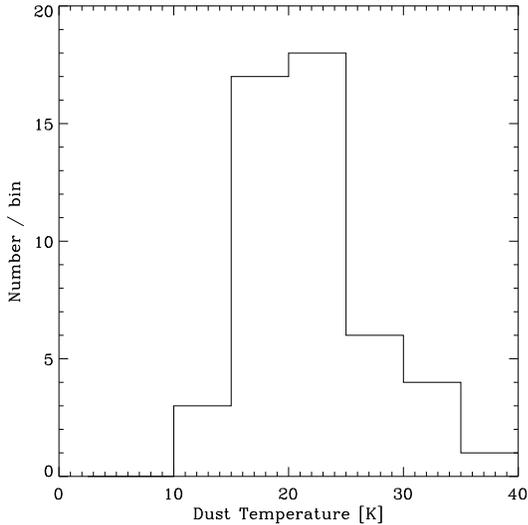}
\caption{Histogram of inferred dust temperatures, assuming a dust
  emissivity index $\beta=1.5$, and excluding 11 objects for which the
  BLAST colors are considered unreliable. The entire distribution
  shifts downward by $\sim5$\,K if instead $\beta=2.0$ is adopted
  (uncertainties are discussed in \S\ref{sec:sedfits}).}
\label{fig:temphist}
\end{figure}

The range of temperatures inferred from the SED fits to BLAST data
with reliable colors is shown in Figure~\ref{fig:temphist}. The survey
has convincingly detected sources with a range of FIR peak
wavelengths, since the temperature span, $10\,\mathrm{K}\la T\la
40\,\mathrm{K}$, is much larger than the individual uncertainties that
are typically $\sim1$--2\,K (Table~\ref{tab:sed}). It should be noted
that these latter uncertainties characterize the range of plausible
temperatures {\it under the assumption of the simple SED model}
(Equation~\ref{eq:sed}). There is also a systematic dependence on
$\beta$; for example, increasing $\beta$ from 1.5 to 2.0 shifts most
temperatures lower by about 5\,K, and derived masses higher by a
factor of $\sim$2. Further details on correlations are given in
\S\ref{sec:sedfits}.

As described above, the luminosity-to-mass ratio can also be
calculated from the SED fits without any knowledge of the distance.
For the simple SED model adopted, $L_{\mathrm{FIR}}/M_\mathrm{c}
\propto T^{4+\beta}$ and so we find a broad range in
$L_{\mathrm{FIR}}/M_\mathrm{c}
\sim0.2$--$130$\,L$_\odot$\,M$_\odot^{-1}$ for the objects with $T$ in
the range 12 -- 40\,K (Figure~\ref{fig:temphist}).

\section{DISTANCES}
\label{sec:dist}

To deduce luminosities and masses one must have an estimate of the
distance to each source. There is, however, little information in the
literature regarding distances to known objects in this field.
Cross-correlation with the {\it IRAS} PSC identifies 23 of the 60
BLAST sources with known objects (Table~\ref{tab:iras}).  While some
of these sources have been studied extensively \citep{beu02,bel06},
few have unambiguous distance information.
Important clues are provided by the morphology and velocity of the
interstellar material and their inter-relationship.
We have carried out a multi-wavelength assessment comparing the BLAST
images and point sources with {\it IRAS} \citep{cao1997,kerton2000};
{\it MSX}; IRAC (\S\ref{sec:mips70}); the STScI Digitized Sky
Survey\footnote{\url{http://archive.stsci.edu/dss/}} (DSS); 21\,cm
radio continuum imaging from the VLA Galactic Plane Survey
\citep[VGPS,][]{stil2006}; and spectral line imaging in
$^{13}$CO\,(1$\rightarrow$0) \citep[FCRAO:][]{brunt2007} and
\ion{H}{1} (VGPS).

\begin{figure}
\centering
\includegraphics[width=\linewidth]{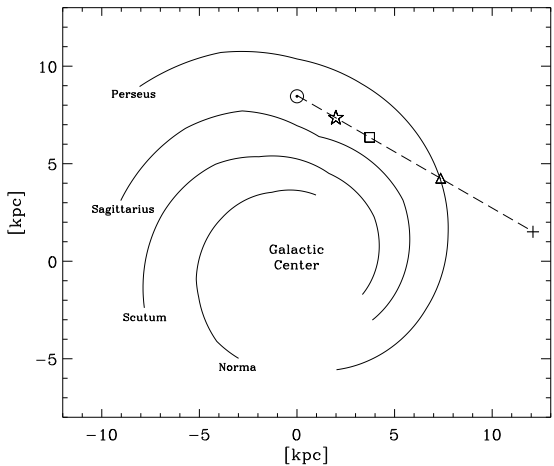}
\caption{Galactic spiral arm model of \citet{Taylor93}.  The dashed
  line indicates the line of sight towards 60$^{\circ}$ longitude.
  The star indicates the location of the open cluster NGC~6823 at the
  adopted distance of 2.3\,kpc \citep{massey1995}.  We associate 49 of
  the 60 BLAST sources with the molecular complex within which
  NGC~6823 has already formed.  Another 10 ``Perseus arm sources'' are
  approximately 8.5\,kpc distant (triangle) and one object (V07)
  appears to be in the outer Galaxy (plus).  The square indicates the
  tangent point. }
\label{fig:spiralarms}
\end{figure}

\begin{figure}
\begin{center}
\includegraphics[width=\linewidth]{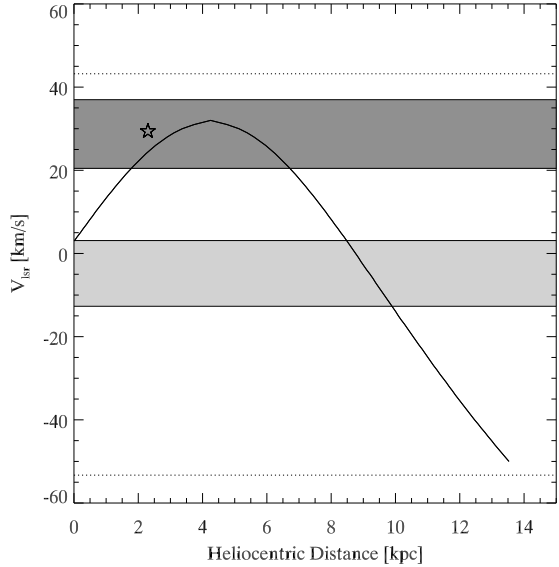}
\end{center}
\caption{The solid curve shows the expected kinematic
  distance-velocity relationship derived from the Galactic rotation
  curve of \citet{bra93} projected along $\ell = 60^{\circ}$.  At the
  adopted distance of 2.3\,kpc for NGC~6823 we show the average radio
  recombination line velocity of Sh2-86 as a star.  The dark shaded
  region indicates the spread of $^{13}$CO velocities measured for the
  BLAST sources associated with the main cloud complex.  The uppermost
  dotted line indicates the velocity of V06, which is higher than the
  main spread but still consistent with systematic motions around the
  mean. The light shaded region indicates the ``Perseus arm sources''
  and the lowest line the source V07 in the outer Galaxy.  }
\label{fig:kinematic}
\end{figure}

We argue below that 49 of the 60 BLAST sources in this field are
associated with the molecular cloud complex within which the open
cluster NGC~6823 has already formed (Figure~\ref{fig:blastmaps}).  We
are then able to adopt the photometric distance measured for the
stars.  We use 2.3\,kpc \citep{massey1995}, though we note that
distances in the range 1.5 to 3.2\,kpc have been reported
\citep{hoyle2003,Guetter92,pena2003}.  
In a program measuring parallaxes using methanol masers
\citep{menten2007} a distance of $2.20 \pm 0.01$\,kpc is found for the
masers associated with V30 \citep[IRAS 19410+2336; see][]{szym2000},
about $0\fdg5$ from the cluster center.
NGC~6823 is placed in a Galactic context in Figures
\ref{fig:spiralarms} and \ref{fig:kinematic}.  In neither position nor
velocity \citep[see below and][Figure~2]{lockman1989} is it part of
the main Sagittarius arm.  In addition ten of the sources appear to be
in the more distant Perseus arm ($\sim8.5$\,kpc), and a single object
(V07) is thought to lie beyond that in the outer galaxy
($\sim14$\,kpc). The sources that are not associated with the NGC~6823
molecular complex are indicated thus in Table~\ref{tab:src}.

The {\it IRAS} images (see Figure~\ref{fig:coverage}) show that the
bright sources and diffuse emission represent a distinctive
enhancement of size $\sim 2^\circ$, which as we discuss below is also
the scale of the coherent molecular complex.  However, within the
complex there is sub-structure, both in space and velocity.
The angular diameter of the NGC~6823 cluster is $\sim 0\fdg5$
\citep{Kharchenko05}, and this is a part of the overall complex.  The
diffuse radio and optical emission from the \ion{H}{2} region Sh2-86
(LBN~135), the portion of the complex that is being ionized, is also
of this size.  Within this region there are ionization fronts, clearly
associated with the ionizing stars in NGC~6823, sculpting the parent
molecular material.
A particularly striking example that can be seen in the red DSS image
is a silhouetted ``pillar'' pointing at the stars, with an ionization
front at its end and V39 immediately behind it. V43 and V49 are other
similar examples.

At 2.3\,kpc, $\sim 2^\circ$ corresponds to a physical size of 80\,pc,
so that all of the parts of the complex are at essentially the same
distance to within a few percent (there could be a systematic error in
the average distance, as noted), with variations indicated by slightly
different velocities or a silhouette indicating relative position
along the line of sight.

To introduce the velocity scale, the radio recombination line velocity
of Sh2-86 measured by \citet{lockman1989} in a 9\arcmin\ beam is $29.4
\pm 2.1$\,\kms\ with FWHM $24.2 \pm 2.9$\,\kms.  This velocity width,
greater than a typical thermal width, and typical of widths found in
that survey, could reflect accelerated gas motions in the ionized
material and/or be intrinsic to the original neutral components
comprising this star forming complex. It is certainly unrelated to
differential Galactic rotation.  For a distance of 2\,kpc, the
gradient in the rotation curve (Figure~\ref{fig:kinematic}) is $\sim
125$\,pc\,km$^{-1}$\,s, and so the velocity spread would in that
interpretation imply a cloud of extreme elongation along the line of
sight.  Note that this velocity and distance combination does not fall
on the mean rotation curve.  There are substantial systematic motions
in this direction, including considerable emission of both $^{13}$CO
and \ion{H}{1} beyond the nominal tangent-point velocity ($\sim
34$\,\kms) and so kinematic distances from the rotation curve in this
range are unreliable.

\subsection{Velocities}
\label{subsec:velocity}

\begin{figure}
\centering
\includegraphics[width=3in]{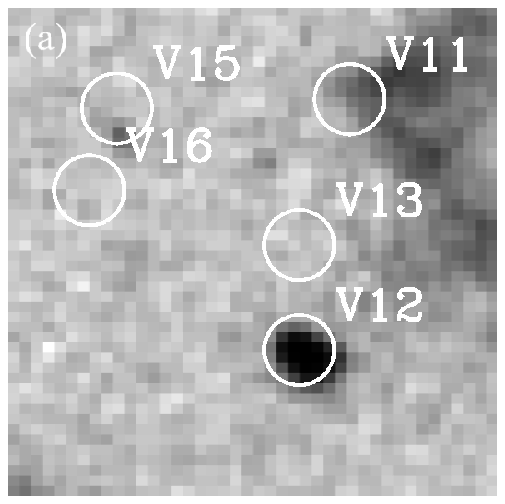}
\includegraphics[width=3in]{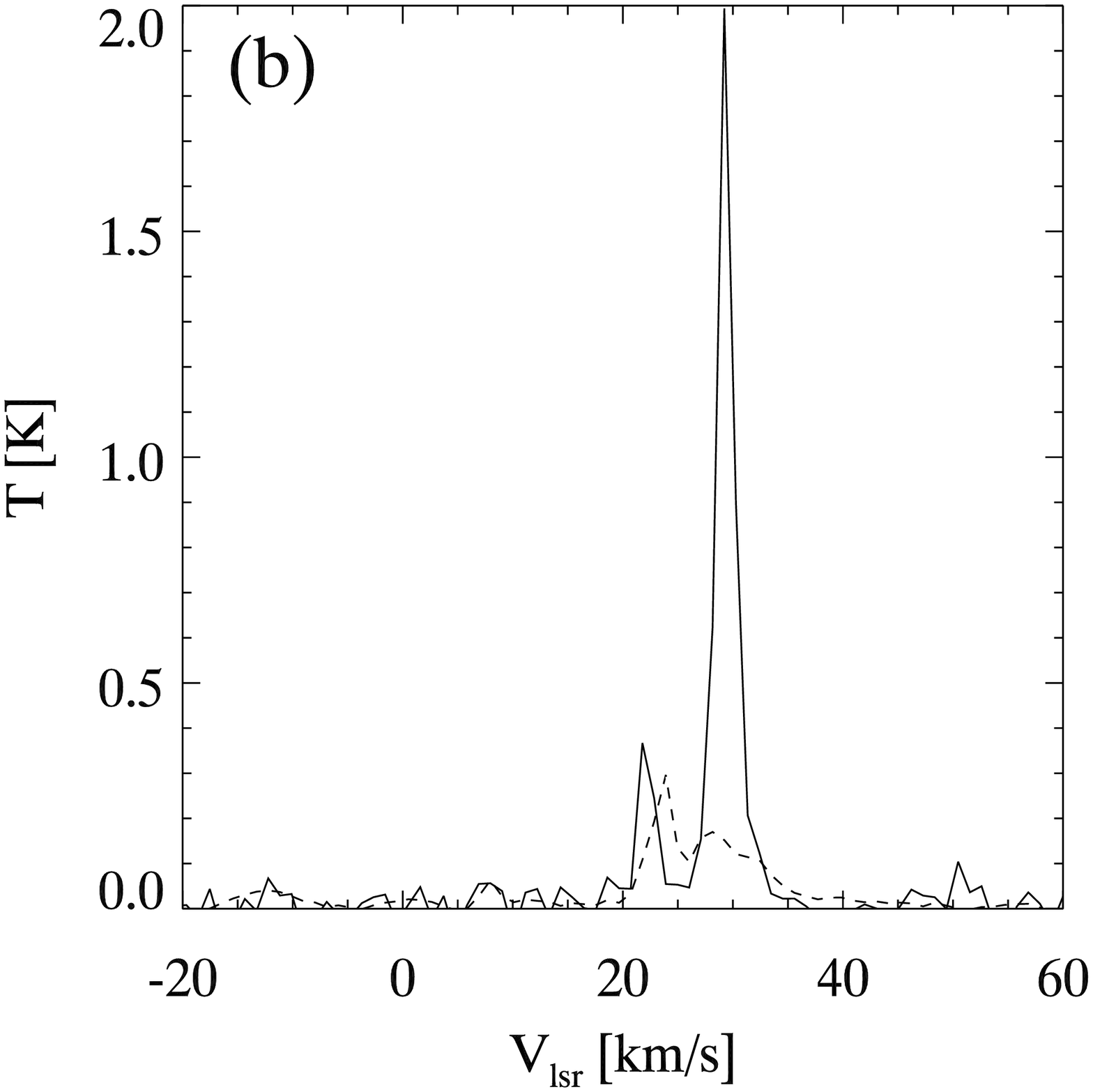}
\caption{({\it a}) Channel image of $^{13}$CO at a velocity of
  29\,\kms\ for the region shown in Fig.~6.  ({\it b}) Plot of
  $^{13}$CO spectrum averaged over 9 pixels centered on V12 (solid
  line) and average over whole region in panel a (dashed).  The bright
  clump is compact (has a high contrast) both spatially and in
  velocity, making it reasonable to identify it with the BLAST source
  (Table~2).  }
\label{fig:COfig}
\end{figure}

\begin{figure}
\begin{center}
\includegraphics[width=\linewidth]{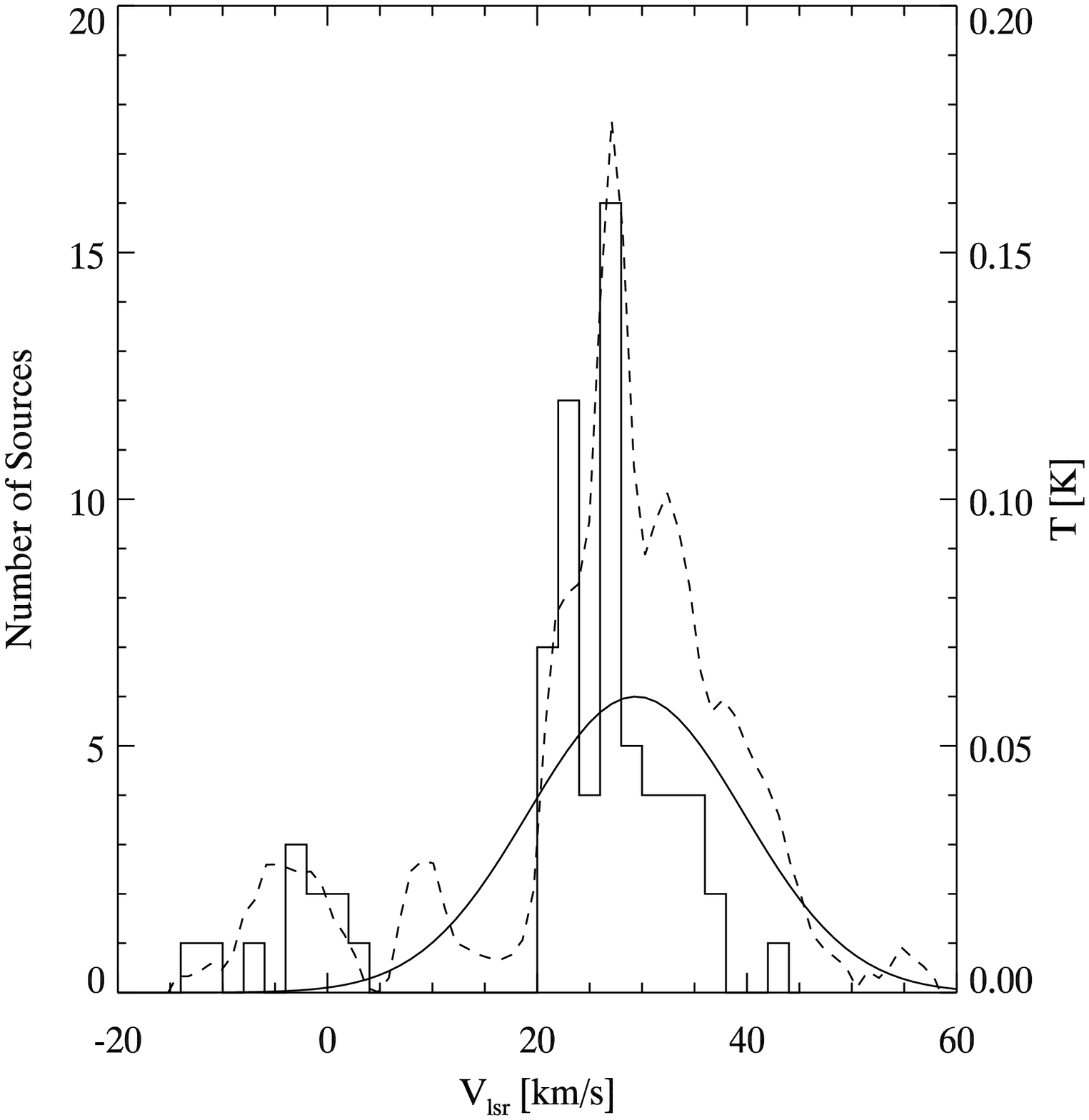}
\end{center}
\caption{Histogram of the $^{13}$CO radial velocities associated with
  BLAST sources.  The majority of these velocities are very similar to
  the radio recombination line velocity of the \ion{H}{2} region
  Sh2-86 measured by \citet{lockman1989}, shown as the Gaussian (with
  arbitrary normalization).  The dashed line indicates the average
  spectrum of the entire $^{13}$CO map, including more diffuse gas.
  The peak at $10$\,\kms\ is the fairly structureless local gas, and
  that at $-5$\,\kms\ is the Perseus arm in which we identify 10
  sources.
}
\label{fig:velhist}
\end{figure}

Using the accurate BLAST source positions, we have examined
$^{13}$CO\,(1$\rightarrow$0) spectra of Vulpecula, having 46\arcsec\
spatial and 1\,\kms\ velocity resolution (binned from a native
resolution of 0.13\,\kms), respectively. Generally a single compact
spatial/spectral coincident feature is identified (see the example of
source V12 in Figure~\ref{fig:COfig}).  We therefore fit Gaussian
profiles to the velocity component at the position of each source.
The lines at BLAST positions have a typical FWHM of 2.5\,\kms.  Ten of
the BLAST/{\it IRAS} counterparts have previously-measured velocities
in other spectral lines, specifically CS \citep{beu02,bro96} and
NH$_3$ \citep{mol96,Sridharan02,zinchenko1997}.  Our measured
velocities for these objects are on average within 1\,\kms\ of the
published velocities, the maximum deviation being 2\,\kms.  The
velocities are listed in Table~\ref{tab:src} and are presented in
Figure~\ref{fig:velhist}.  The broad peak from 21 to 36\,\kms\ is
similar to that seen in the average spectrum of this region, which
also includes more diffuse gas.  The $^{13}$CO velocity of V06, at
43\,\kms, still falls well within this range.

\begin{figure} 
\centering
\includegraphics[width=\linewidth]{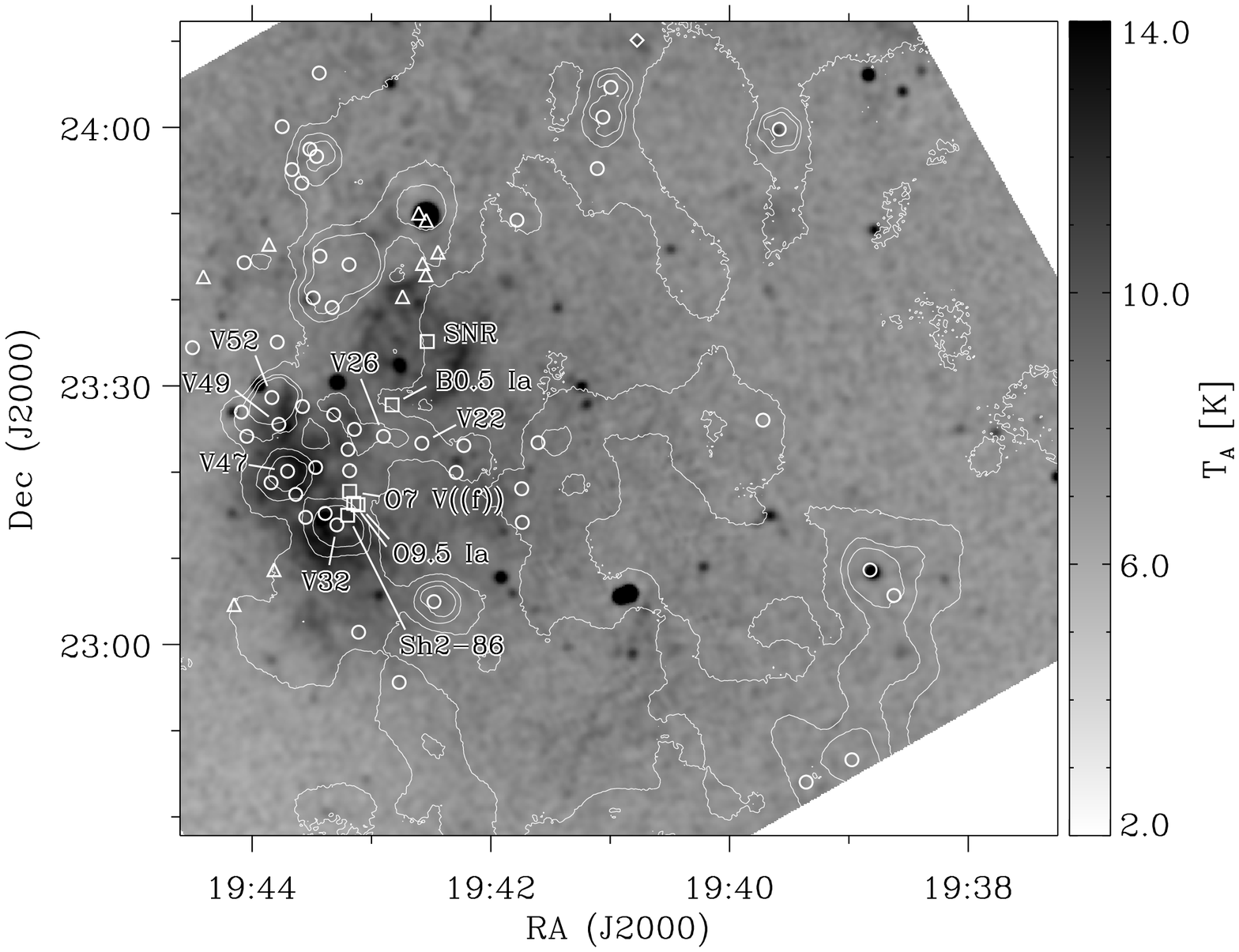}
\caption{The grayscale shows the VGPS 21-cm radio continuum emission,
  which dominated by a structured \ion{H}{2} region Sh2-86 and a shell
  supernova remnant \citep[diameter $\sim15'$,][]{taylor1992}.
  Overplotted are the BLAST 350\,\micron\ contours, which reveal the
  interfaces between thermal dust emission (neutral material) and the
  ionized gas.  Circles (sources associated with NGC~6823) and
  triangles (Perseus arm sources) identify the compact BLAST sources.
  The squares indicate the most massive stars in the open cluster
  NGC~6823. Labeled BLAST sources (V22, V26, V32, V47, V49 and V52)
  lie in molecular clumps that clearly affect the radio emission from
  Sh2-86. }
\label{fig:radio}
\end{figure}

Several BLAST sources are in molecular clumps that clearly affect the
radio emission from the extended \ion{H}{2} region Sh2-86 seen in
Figure~\ref{fig:radio}. For example, V32 is a clear local minimum
surrounded by an ionization front. The pair V22 and V26 are also at a
minimum.  Other objects apparently shaping or influencing the
steepness of the radio contours are V47, V49, and V52.  All of these
sources have CO clump velocities in the range 26--31\,km\,s$^{-1}$,
like the radio recombination line velocity.

V23 (with V19 close by) is coincident with a bright compact \ion{H}{2}
region and has the weakest CO signature of all the BLAST sources.
Additional velocity information for V23 was obtained using \ion{H}{1}
absorption against the radio continuum source using VGPS data.
Velocity components are identified at $-5$, 7.5, 21 (very strong), 27,
and 31\,\kms.  The radio recombination line velocity in a 3\arcmin\
beam \citep{lockman1989} is $-2.8 \pm 1.8$\,\kms\ with FWHM $15.2 \pm
2.6$\,\kms (the unusually low line width suggests an electron
temperature less than 5000\,K).  The spatial position of V23 with
respect to the main CO complex and other {\it IRAS} and BLAST sources
suggest it is at the same distance, but the velocities at $-5$ and
$-2.8$\,\kms\ indicate a much larger distance $\sim9$\,kpc just beyond
the solar circle in the Perseus arm.
Nearby, V20 and V24 have similarly low velocities, appearing along a
clumpy arc of $^{13}$CO emission at $-3$ to +2\,\kms.  V17 and V21,
which have $^{13}$CO components near $\sim0$\,\kms\ that are
comparable in strength to weak components in the nominal NGC~6823
molecular cloud velocity range, are probably in this distant cloud
too.  We shall call these the ``Perseus arm sources.''

The only other BLAST source for which we can measure the \ion{H}{1}
absorption directly is V02.  It has strong sharp absorption at 31 and
35\,\kms\ and weaker components at 10\,\kms, and possibly $-6$ and
+43\,\kms as well.  The $^{13}$CO velocity of this clump is 32\,\kms,
but there is considerable gas near $\sim$40\,\kms\ in the surrounding
cloud on this side of the map.

There is a weak radio source that is coincident with V07 and a
molecular clump at $-54$\,\kms. This source is too faint in the radio
for any possibility of measuring \ion{H}{1} absorption.  We use the
rotation curve (Figure~\ref{fig:kinematic}) to find a kinematic
distance of 14\,kpc, making this an intrinsically luminous star
forming region in the outer Galaxy.

On the low side of the velocity distribution, there are a few other
sources with $^{13}$CO components near $\sim0$\,\kms\ that are
comparable in strength to weak components in the nominal NGC~6823
molecular cloud velocity range, namely V51, V54, V58, and V59.
Using C$^{18}$O spectra \citep{brunt2007}, which probe even denser
gas, the low velocity component is chosen for V51 and V58 (these join
the Perseus arm sources).
V54 and V59 are seen in projection near the lower outskirts of the
NGC~6823 cloud, as depicted in the zeroth-moment image
(Figure~\ref{fig:co0blast}). Since there is no compelling
morphological evidence to relate them to the higher-velocity gas, we
also consider them to be Perseus arm sources.

Finally, we note a puzzling result.  Using the \ion{H}{1} absorption
technique to break the distance ambiguity, and Arecibo observations
(with a 4\arcmin\ beam) at the Sh2-86 position \citep{lockman1989},
\citet{kuchar1994} find absorption in the spectrum out to 50\,\kms and
so adopt the far distance solution 6.3~kpc.  Certainly, there is no
question about the velocity of the \ion{H}{2} region and hence the
associated molecular cloud and BLAST sources.  But as we have
mentioned above, the ionization fronts seen on the DSS protruding into
the molecular cloud are clearly caused by the bright OB stars in
NCG~6823 identified by \citet{massey1995}, and these have a
photometric distance much closer to the near solution.  We have looked
at the VGPS \ion{H}{1} data (also with continuum, and 1\arcmin\
spatial resolution) across the face of the radio nebula.  We find that
there are large changes in the emission and absorption spectrum, with
some absorption to 50\, \kms (in line wings) even at the fainter
western edge, where the continuum emission blends into the (non-zero)
background.  This suggests that there is variable distant background
continuum emission that is making it difficult to apply the technique
on this extended source.

\subsection{Morphology}
\label{subsec:morphology}

\begin{figure} 
\centering
\includegraphics[width=\linewidth]{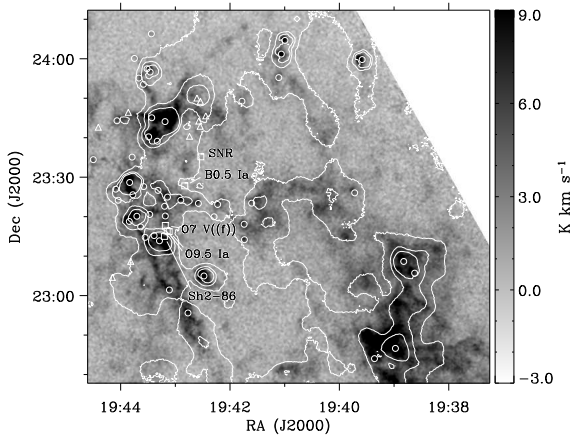}
\caption{The grayscale shows the $^{13}$CO emission integrated over 21
  to 36\,\kms.  Overplotted are the BLAST 350\,\micron\ contours which
  demonstrate a close correspondence between thermal dust emission and
  molecular gas in this velocity range.  Circles (sources associated
  with NGC~6823) and triangles (Perseus arm sources) identify the
  compact BLAST sources.  The squares indicate the most massive stars
  in the open cluster NGC~6823.  Also indicated is the position of the
  supernova remnant, and the extended \ion{H}{2} region Sh2-86.  }
\label{fig:co0blast}
\end{figure}

\begin{figure}
\centering
\includegraphics[width=\linewidth]{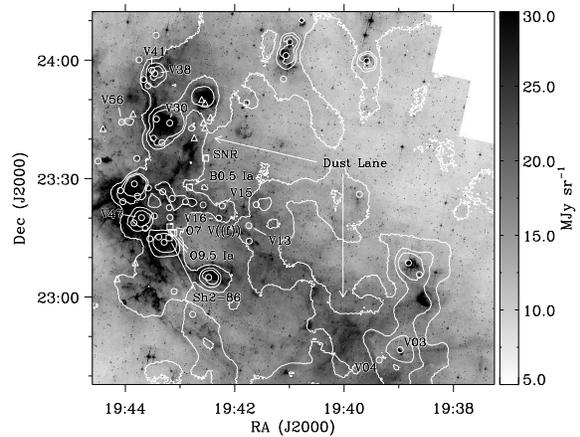}
\caption{The grayscale shows IRAC 8\,\micron\ emission, with the same
  annotations as Figure~\ref{fig:co0blast}.  We note that the labeled
  BLAST sources (V03, V04, V13, and V56) occur at minima in the
  diffuse 8\,\micron\ emission, possibly indicating that the
  submillimeter objects are (associated with) infra-red dark clouds
  somewhat in the foreground. A prominent filament which appears as 
  a dust lane in the red DSS image is also indicated (probably related to
  the BLAST sources V13, V15, V16, V38, and V41). Optical absorption is
  also associated with V47. These features are 
  discussed in \S\ref{subsec:morphology}.
  \label{fig:iracblast}}
\end{figure}

A map of the $^{13}$CO emission integrated over the velocity interval
21--36\,\kms\ is shown in Figure~\ref{fig:co0blast}.  
Figure~\ref{fig:iracblast} shows the IRAC 8\,\micron\ image for
comparison.  These both trace neutral material, but the latter,
heavily influenced by PAH emission, requires illumination by
sub-ionizing ultraviolet radiation (rather than simply tracing column
density).  Also, 8\,\micron\ is a short enough wavelength for there to
be extinction by large column densities, producing the infra-red dark
cloud (IRDC) phenomenon \citep{ega98,simon2006}.
A portion of the molecular cloud associated with V03 and V04 produces
a local minimum in the diffuse 8\,\micron\ emission, suggesting it is
in the foreground.  On a smaller scale, V13 and V56 appear to exhibit
similar behavior.

Overlaid on the figures are the BLAST 350\,\micron\ contours, which
reveal excellent correspondence between this measure of the neutral
column density and the other tracers.  Also indicated are the BLAST
point sources and the ionizing stars in NGC~6823 \citep{massey1995}.

A particularly striking filament, which is a prominent dust lane in
the red DSS image, is seen as a ridge in the BLAST, $^{13}$CO, and
IRAC maps (indicated in Figure~\ref{fig:iracblast}), running from V30
to the south west (parallel to the Galactic Plane) toward V03.  BLAST
sources V13, V15, V38, V41, and possibly V16 appear to be related to
this optically-dark filament.  Other optical absorption is associated
with BLAST source regions, such as V47 projected on the \ion{H}{2}
region.

An interesting phenomenon seen in the \ion{H}{1} data is
self-absorption (HISA), caused by colder foreground material; it is
most convincingly found using the criteria of narrow lines and
fine-scale spatial structure \citep{Gibson2000} but can also be more
widespread.  HISA appears at 24\,\kms\ in this filamentary structure.
Near 33\,\kms\ there is striking filamentary HISA which extends into
the V22--V26--V28 portion of the molecular complex, where the lack of
\ion{H}{1} emission no doubt also reflects a real deficit. At
36\,\kms\ there is more HISA which traces the diffuse submillimeter
emission along the ridge ending at V06.  This ridge is evident in
$^{13}$CO at these velocities, but is not so clearly seen with IRAC.

Examination of the sequence of channel maps in the $^{13}$CO data
cube, or the first moment (average velocity) map reveals a velocity
gradient from the north-east to the south-west (higher to lower
Galactic longitudes).  Along the long filament the velocity changes
quite systematically, indicating a large scale connectivity between
the different portions of the molecular complex.

Also revealed in these data are large areas of low emission.
To the west of the stellar cluster is a distinctive void, cleared of
molecular gas and dust.
As expected, because the gas has become molecular, there is a general
deficit of \ion{H}{1} across this entire region. This can be seen
extending to adjacent longitudes in the longitude-velocity diagram of
\citet{stil2006}.  The deficit is exaggerated adjacent to the
\ion{H}{2} region (near 33\,\kms\ where the gas is molecular).  This
can appear more localized too, e.g., near V03 and in the V08--V09
cloud at the same velocity as the $^{13}$CO.

Another void coincides with a supernova remnant found by
\citet{taylor1992} and is seen in finer detail in the VGPS radio
continuum image (Figure~\ref{fig:radio}).  The void is seen in
\ion{H}{1} and in $^{13}$CO as well near 22\,\kms.  The remnant is not
noticeably interacting with the surrounding dense material (e.g., no
steepening of the $^{13}$CO contours or brightening or deformation of
the radio shell), and so perhaps the supernova occured in a
pre-existing cavity, rather than creating it.  The diameter of the
remnant is only 0\fdg2 (8\,pc) and so it could be at the average
distance of this cloud complex (as the velocity suggests) but not yet
have encountered one of its clumpy components.

\section{DUST MASSES AND LUMINOSITIES}
\label{sec:mass}

In \S\ref{sec:dist} we determined distances for all of the objects: 49
are associated with the open cluster NGC~6823 at $\sim2.3$\,kpc, 10
with the Perseus arm at $\sim8.5$\,kpc, and 1 object is in the outer
galaxy at $\sim14$\,kpc (summarized in Table~\ref{tab:src}). Here we
use these distance estimates to convert the integrated fluxes from
\S\ref{sec:coldsed} into bolometric luminosities, $L$, and to derive
the mass, $M_\mathrm{c}$, for each source.

The clump mass $M_\mathrm{c}$ given in column 3 of Table~\ref{tab:sed}
has been obtained from Equation~\ref{eq:mass}.  Note that although a
mass absorption coefficient at 250\,\micron\ appears in this
expression, $A$ is derived from the fit to the entire submillimeter
and FIR data, rather than just the single color-corrected BLAST
250\,\micron\ data point alone.

The masses of the clumps discovered range over
$\sim15$--700\,M$_\odot$ in NGC~6823, extending to somewhat higher
masses in the Perseus arm. These sources are by definition compact in
the submillimeter (and they are compact in CO as well), with
characteristic sizes $D < 0.44$\,pc at 2.3\,kpc (taking the
de-convolved 250\,\micron\ map beam size of 40\arcsec\ from
Table~\ref{tab:mapsens} as a limit). If observed at higher resolution
many of these ``clumps'' would probably qualify as ``cores'' in the
terminology of \citet{zinn2007}.  The column density for a typical
100\,M$_\odot$ clump, assuming a uniform sphere limiting diameter $D$,
is of order $10^{23}(0.44\,\mathrm{pc}/D)^2$\,cm$^{-2}$ (visual
extinction to the cloud center $A_V \sim 22(0.44\mathrm{pc}/D)^2$ and
density of order $10^5(0.44\,\mathrm{pc}/D)^3$\,cm$^{-3}$).  The
Bonner-Ebert critical mass for a gravitationally-bound clump, above
which collapse of an unstable configuration occurs, is given by
$M_{\mathrm{crit}}/M_\odot \approx (D/\mathrm{pc})
(T_{\mathrm{gas}}/\mathrm{K})$ \citep[see
e.g.,][Equation~4]{kerton2001}.  In these dense clumps,
$T_{\mathrm{gas}}$ is probably close to the dust temperature and so
for the derived $T$ and $M_{\mathrm{c}}$ most of the clumps that we
have identified would be unstable and are likely supported by
additional pressure from turbulence or magnetic fields.  On release of
this pressure, they have the potential to form a massive star or even
a small cluster.  Interestingly, several of the lowest mass clumps
would be just stable if they were as extended as 1\,pc.

The range of luminosities does indeed extend to $10^4$\,L$_\odot$ in
NGC~6823, which corresponds to the luminosity of a B0.5 zero age main
sequence star \citep{panagia1973}.  But as we shall discuss further,
at the low end, $\sim40$\,L$_\odot$, there is by contrast no
indication of significant star formation.

\begin{figure}
\centering
\includegraphics[width=\linewidth]{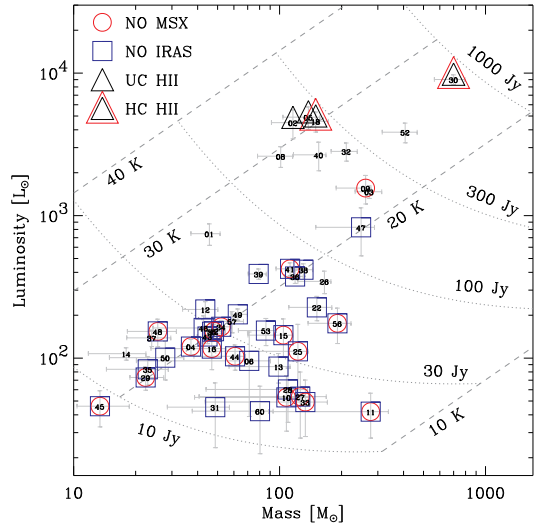}
\caption{Luminosity vs.\ mass for the 49 objects believed to be
  associated with NGC~6823.  Uncertainties in mass and luminosity are
  estimated from the range of temperatures consistent with the BLAST
  and FIR data under the assumption of the simple SED model of
  Equation~\ref{eq:sed}.  Not included are additional sources of error
  involving the distances to the objects and adopted values of
  $\beta$, $\kappa_0$, and gas-to-dust mass ratio.  The dashed lines
  are loci at fixed temperatures (and $L_{\mathrm{FIR}}/M_\mathrm{c}$)
  of 10--40\,K, assuming a modified blackbody SED with $\beta=1.5$.
  Orthogonal to these are loci (dotted lines) of constant 250\,$\mu$m
  flux density, ranging from 10 to 1000\,Jy, using the same model.
  Sources lacking {\it IRAS} PSC or {\it MSX} counterparts are marked
  with circles and squares respectively. Ultra Compact \ion{H}{2}
  region and Hyper Compact \ion{H}{2} region candidates are indicated
  with single and double triangles, respectively.}
\label{fig:lumvsmass}
\end{figure}

\begin{figure}[h]
\centering
\includegraphics[width=\linewidth]{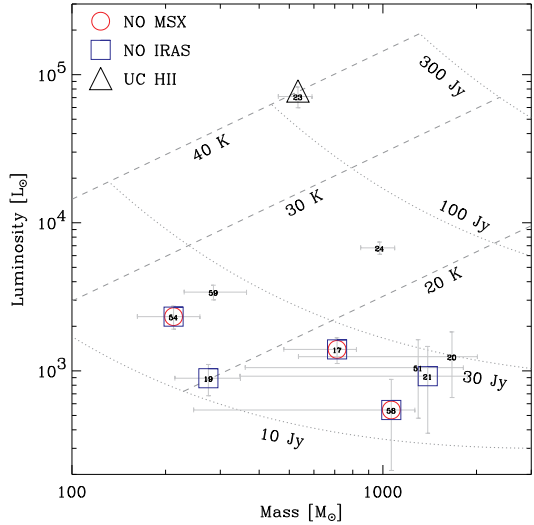}
\caption{Same annotations as Figure~\ref{fig:lumvsmass}, but for the
  10 objects that are associated with the Perseus arm.  }
\label{fig:lumvsmassPerseus}
\end{figure}

Luminosity as a function of mass is shown in
Figures~\ref{fig:lumvsmass} and \ref{fig:lumvsmassPerseus}. This is
essentially a reparameterization of the BLAST flux density vs.\
temperature plane, in terms of intrinsic source parameters.  Because
of the different distances for the two main groups, the Perseus arm
sources and those associated with NGC~6823, the scaling of the axes is
different, requiring similar but separate diagrams.
Recall that the derived quantities are based on a SED fit assuming a
modified blackbody SED with $\beta=1.5$.  Using this same model, the
dotted lines are loci (with varying $T$) of constant 250\,$\mu$m flux
densities ranging from 10 to 1000\,Jy.
Orthogonal to these, the dashed lines are loci (with varying observed
flux density) at constant $T=10$, 20, 30, and 40\,K.  With our adopted
model, these correspond to constant $L_{\mathrm{FIR}}/M_\mathrm{c}$,
values of 0.07, 3.2, 30, and 140\,L$_\odot$\,M$_\odot^{-1}$,
respectively.
The 11 objects with unreliable BLAST colors
(\S\ref{sec:sourceid}, marked in Table~\ref{tab:src}) have been
assigned a temperature of 20\,K and so are seen distinctly along that
locus.  If these sources have different temperatures (probably lower
given the usual lack of {\it IRAS} PSC counterparts), they would move
along the constant flux density locus (down and right if cooler).
These Figures can be used to investigate which early stages of star
formation might be present.  To this end, sources have been marked
which lack {\it IRAS} PSC or {\it MSX} counterparts, or are UC
\ion{H}{2} regions

\subsection{Mass Spectrum} \label{sec:spec}

\begin{figure}
\includegraphics[width=\linewidth]{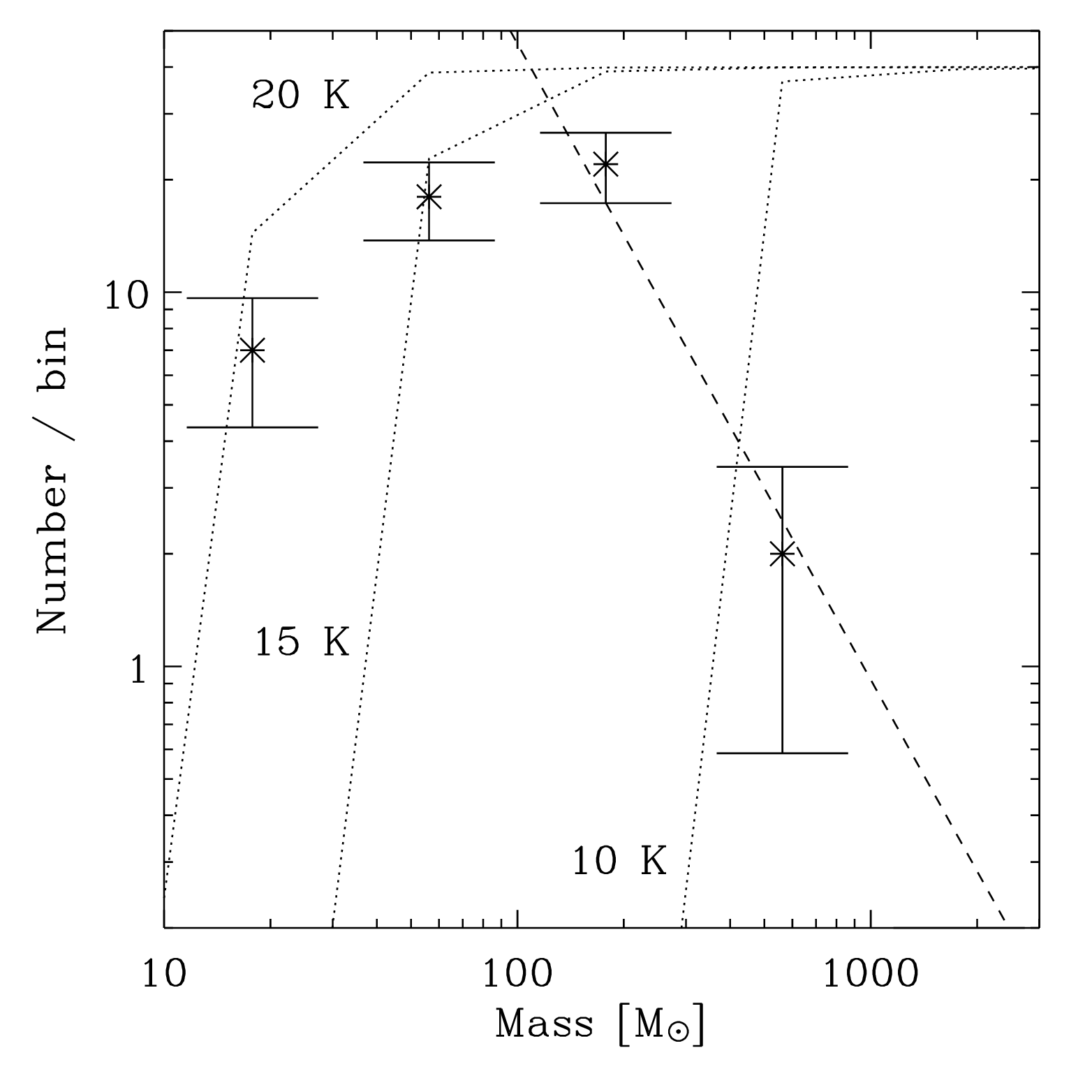}
\caption{Mass function of the 49 objects associated with
  NGC~6823. Poisson error bars are shown, although there are
  additional sources of uncertainties for the parameters of
  Equation~\ref{eq:mass} for each object.  For reference, the dashed
  line is a power-law, $150(M/50\,\mathrm{M}_\odot)^{-1.7}$.
  The two lowest bins suggest a turnover, but this shape is probably
  due to incompleteness.  To illustrate this the 250\,\micron\ flux
  density completeness (Figure~\ref{fig:blastcomplsims}) has been
  converted to a mass completeness using Equation~\ref{eq:mass}
  assuming various temperatures $T=10$, 15, and 20\,K.  These are
  shown as dotted lines, arbitrarily scaled such that 100\%
  completeness is 40 objects per bin.}
\label{fig:mfunc}
\end{figure}

The clumps that we have detected are much more massive than stars and
therefore represent an early stage in star formation that is not
easily, even empirically, related to the stellar initial mass
function.  Fragmentation and efficiency affect the transfer function
in unknown ways.  Nevertheless, it is interesting to investigate the
mass spectrum, which we have done for the 49 BLAST sources associated
with the NGC~6823 cloud complex.  The masses of individual sources
were placed in logarithmically spaced bins (2 per decade); a lower
limit on the error is then estimated from the Poisson uncertainty for
each bin.  The resulting mass function is shown in
Figure~\ref{fig:mfunc}.

At masses $\ga100$\,M$_\odot$, the distribution is consistent with the
mass functions of molecular clouds. For example, \citet[and references
therein]{kra98} found a power-law index $\alpha \simeq 1.7$ in a mass
range which encompasses the mass interval considered here.  For
reference, we have shown their power-law scaled vertically such that
it passes through the high-mass bins in Figure~\ref{fig:mfunc}: $dN/dM
= 150(M/50\,\mathrm{M}_\odot)^{-1.7}$.  Note that adopting a different
value of dust emissivity affects the inferred dust mass, since it is
highly correlated with the fitted temperature (\S\ref{sec:sedfits}),
and hence mass. For example, choosing $\beta=2.0$ decreases
temperatures by $\sim5$\,K, and increases cloud masses by a factor
$\sim2$. This effect shifts the mass function to the right in this
figure, but does not affect the power-law index.

We note that the mass function in our sample appears to turn over at
masses below $\sim100$\,M$_\odot$. However, the completeness of the
survey also drops at these masses, with uncertainties that are
difficult to quantify given the range of object temperatures sampled
(Figure~\ref{fig:lumvsmass}).
To attempt to assess the mass completeness, we converted the $x$-axis
of the flux density completeness function
(Figure~\ref{fig:blastcomplsims}) into mass using
Equation~\ref{eq:mass}, assuming consistently a distance of 2.3\,kpc
and $\kappa_0 = 10$\,cm$^2$g$^{-1}$.  Some results are shown in
Figure~\ref{fig:mfunc}.  If we assumed that the sources potentially
missed had a temperature of $20$\,K, then completeness has a
negligible effect, but for lower temperatures, which seem more
probable, the effect is significant, because lower temperatures
correspond to higher masses for a given flux density.  See
\citet{johnstone2000} and \citet{enoch2006} for further discussions
regarding the completeness of mass spectrum estimates from
submillimeter surveys.
If, however, the turnover were real at $\sim100$\,M$_\odot$, our
survey may be probing an interesting effect discussed by other authors
\citep[see][and references therein]{bel06}.  They claim that the point
at which the mass distribution flattens is proportional to the mass
range of clumps in star forming clouds.

\section{DISCUSSION}  \label{sec:disc}

\subsection{Stages of Pre-stellar Evolution}  
\label{sec:prestellar}

This particular survey was not unbiased, since it targeted a region
which has already formed massive stars and which contains bright {\it
  IRAS} sources which are protostar candidates.  But it does fulfill
one of the main goals of BLAST, to explore the sites of formation of
high-mass stars and to examine the pre-stellar evolutionary sequence.
In particular, with the BLAST wavebands we are sensitive to the
earliest phases of massive star evolution, which would be molecular
clumps lacking an internal energy source and thus cool with low
$L_{\mathrm{FIR}}/M_\mathrm{c}$.  In Figure~\ref{fig:lumvsmass} there
are indeed many such low luminosity-to-mass ratio sources that are too
faint and cool to have {\it IRAS} and {\it MSX} counterparts.

An interesting perspective is that the luminosity-to-mass ratio for
diffuse cirrus in the local interstellar radiation field is of order
1\,L$_\odot$\,M$_\odot^{-1}$.  The low ratio (and lower temperature)
found in these molecular clumps, even in the presence of a stronger
impinging radiation field, is a result of the tremendous opacity to
this external radiation, uncompensated by an internal source.  In
absolute terms the luminosity of a cool clump can be quite impressive,
because of the large mass, and yet not be a reliable indicator of
there being star formation within.

Once a high-mass pre- or proto-stellar object begins to form, there is
considerable luminosity to be reprocessed, resulting in higher
$L_{\mathrm{FIR}}/M_\mathrm{c}$ and warmer dust.  BLAST can of course
see such objects as well, even those without MIR associations.  There
is not yet a unique definition or set of physical properties
associated with HMPOs, but there is general agreement that they are
heavily obscured objects (even making detection difficult in the MIR)
that have not yet reached the stage at which the emergent Lyman
continuum is sufficient to ionize the surrounding medium to form an
\ion{H}{2} region, and so they are undetected in the radio continuum
as well.  With spectral energy distributions peaking at wavelengths
$\ga 200$\,\micron\ they can be extremely weak and undetectable by
{\it IRAS} at 25\,\micron, thus excluding them from previous surveys
\citep[e.g.,][]{Sridharan02}.  Therefore the most unique BLAST
objects, which are good HMPO candidates, are those with no {\it IRAS}
or {\it MSX} counterparts, and no radio emission.

In the subsequent phase, when the massive protostar has become hotter,
near the zero age main sequence, the ionizing capability is enhanced
and a small region in the dense surroundings is ionized.
Historically, such ``ultra compact'' (UC) \ion{H}{2} regions have been
the primary means to investigate the early phases of OB stellar
evolution.  This is a productive approach, since it deals with
intrinsically luminous objects which are readily detectable, and the
radio emission provides unequivocal evidence for an ionizing star.
The more evolved, and therefore warmer, sources are more likely to
have a MIR counterpart and be detectable by {\it MSX} and/or IRAC,
than the colder and/or more deeply embedded sources.  We do see such
objects, as identified in Figures~\ref{fig:lumvsmass} and
\ref{fig:lumvsmassPerseus} (see \S\ref{sec:uch2}).

In the past decade it has become clear that ``hot-cores'' represent a
stage in the massive star formation process, probably somewhat earlier
than UC \ion{H}{2} regions \citep[see, e.g.,][and references
therein]{kurtz2000}.  The precise details of the transition are still
lacking, and our understanding is complicated by the co-existence of
objects at different evolutionary phases within the same cluster.
Also, the distinction between hot-cores and UC \ion{H}{2} regions is
often dependent on the sensitivity of the available observations, as
several hot-cores are found to emit radio continuum emission if
observed with enough sensitivity and angular resolution (opacity can
also be a factor for dense objects).  Likewise, although some authors
list hot-cores as a subset of HMPOs \citep[e.g.,][]{Sridharan02},
hot-cores are generally warmer, with $T_{\rm k} \ga 100$\,K, whereas
objects called HMPOs usually have temperatures much lower than 100\,K
\citep[e.g.,][]{mol96}.  Also as noted by \cite{Sridharan02}, among
higher luminosity objects, a low luminosity-to-mass ratio relative to
that for UC \ion{H}{2} regions might indicate a HMPO, though
\citet{bel06} found no evidence to support this difference between
their {\it ``Low''} and {\it ``High''} (less and more evolved)
sources.

None of the objects that we detect has an inferred temperature greater
than 40\,K (assuming $\beta=1.5$), even those that are UC \ion{H}{2}
regions.  The temperature that we derive from the SED characterizes
the bulk of the dust emission (more than 50~\%) but of course is a
simplification.  When a protostar forms, the dust nearby is hotter
than in the parent molecular clump, and also appears as a more compact
source.  Thus what is measured and interpreted as mass and luminosity
depends on the radiative transfer and temperature gradient as well as
the angular resolution of the measurements. While $L$ reprocessed by
dust (and the amount of ionization gauged from the radio continuum
emission) are calorimeters for the embedded stellar population that is
forming, the mass is that of the entire parent clump, depleted by
accretion onto stars and later cloud dispersal.  The
luminosity-to-mass ratio can therefore be different, even for the same
stellar population, being maximized for optical depth of order unity
in a compact (and therefore warm) circumstellar region.

A plot like Figure~\ref{fig:lumvsmass} could be used to examine the
evolutionary sequence, whether at constant $M_\mathrm{c}$ or
otherwise, determining the lifetimes for each stage from the relative
numbers of objects.  However, this is complicated in practice by
effects like clustering and detection bias and the fact that star
formation does not proceed at precisely the same rate everywhere in a
molecular clump.  Thus without adequate resolution it would be
difficult to discriminate between material associated with an UC
\ion{H}{2} region and another nearby {\it cold} HMPO, for example.
This problem can be remedied by high-resolution and high dynamic range
observations (as anticipated with {\it Herschel} and ALMA); indeed in
some cases interferometric observations have led to the detection of
HMPOs in the same fields as the UC \ion{H}{2} regions \citep[see,
e.g.,][ and references therein]{olmi2003}.
Likewise, objects in Figure~\ref{fig:lumvsmass} with candidate {\it
IRAS} and/or {\it MSX} counterparts, might in fact harbor a HMPO that
is associated, but not coincident, with a MIR source at a later
evolutionary stage in the same cluster.

\subsection{Ultra Compact \ion{H}{2} Regions}
\label{sec:uch2}

We have detected several {\it bona fide} UC \ion{H}{2} regions, by
which we mean compact objects whose bolometric luminosities {\it and}
radio continuum emission are consistent with an embedded massive star
near the zero age main sequence.  Although within our beam there might
be stars forming over a range of masses, the luminosity (and even more
so the ionizing radiation), is dominated by the most massive and
hottest stars with the earliest spectral types.  Therefore, while not
precise, it is still useful to characterize the region using a single
spectral type.  We use luminosities and ionizing rates from
\citet{panagia1973}, which to the accuracy required here are close to
more modern values \citep{schaerer1997}.

These sources can be been identified as UC \ion{H}{2} {\it candidates}
using color criteria based on {\it IRAS} measurements as in the
\citet{wood1989}, \citet{kurtz1994}, \citet{bro96} and
\citet{Watson03} catalogs (qualitatively, these sources are relatively
weak at 12\,\micron). Using the BLAST data we are able to find very
accurate temperatures.
Radio continuum measurements in the survey region at 1.4\,GHz are
available from the NRAO VLA Sky Survey
\citep[NVSS\footnote{http://www.cv.nrao.edu/nvss/},][]{condon1998} and
the VGPS \citep{stil2006}. In addition, we searched the second and
third MIT-Green Bank 5\,GHz ``MG'' surveys \citep{langston1990,
  griffith1990} which, though much shallower than NVSS, can provide a
second flux density to estimate the spectral index between 1.4 and
5\,GHz.  Some additional targeted observations at 4.9\,GHz using
Arecibo are described in \citet{Watson03}.

We will discuss these sources briefly in order of decreasing distance.
V07 is the source in the outer Galaxy, which though fairly luminous
($3 \times 10^4$\,L$_\odot$), is not among the more conspicuous
sources in the maps. It is consistent with excitation (both luminosity
and ionization) by a B0 star (only 20\,M$_\odot$ as compared to
600\,M$_\odot$ for the clump). In the VGPS radio image this source is
extended and in the NVSS it is double.  The second component, to lower
longitude by about 1\arcmin, coincides with a second {\it IRAS} source which
appears at 12 and 25\,\micron.

V23, in the Perseus arm, is the most luminous source in this BLAST
survey ($7\times10^4$\,L$_\odot$; Figure~\ref{fig:lumvsmassPerseus}),
and a prominent radio, submillimeter, and infra-red source, despite
its distance.  Its excitation is consistent with a star as early as
O7--8 (30\,M$_\odot$, as compared to 500\,M$_\odot$ for the clump).
The emission is extended, overlapping a second BLAST source V19, and
is no doubt more complex than this simple description.

V02 and V05 are among the sources to which we have assigned the
NGC~6823 distance (Figure~\ref{fig:lumvsmass}).  The luminosity
implies a B2 zero age main sequence star, but the ionization (50\,mJy
radio continuum) requires B0.5 (for these relatively cool stars the
ionization is extremely sensitive to the temperature).  A further note
is that the radio source, while being aligned with the {\it IRAS}
source, is offset by 30\,\arcsec\ from V02.

V30, the brightest BLAST source and the most luminous source at this
distance, is by contrast not an NVSS/VGPS source at 1.4\,GHz. However,
using the VLA B array at 8.2\,GHz, \citet{Sridharan02} found weak
emission at the level of 1\,mJy. Lower resolution Arecibo measurements
at 4.9\,GHz \citep{Watson03} give 3\,mJy. Relative to V02 and V05, V30
is under-luminous in the radio continuum by an order of magnitude.
This could be because the central object, though luminous, has not
become hot enough on approach to the zero age main sequence (it is a
HMPO), or because of opacity if the surroundings of the protostar are
very dense.  We note that about 10\% of the sources observed by
\citet{wood1989} are optically thick at 2\,cm, and a higher percentage
may be applicable for the BLAST sources, since they probe the earliest
stages of massive star formation. V30 might be what is called a Hyper
Compact (HC) \ion{H}{2} region \citep[e.g.][]{keto2007}, and has been
marked accordingly in Figure~\ref{fig:lumvsmass}.

The situation for the other comparably-luminous sources V08, V18, V32,
and V52 is less clear.  All but V18 meet the color criteria for being
an UC \ion{H}{2} region. However, none are NVSS sources.  The
detection of the V18 and V30 in targeted Arecibo observations
\citep{Watson03} appears to be contradicted by the lower upper limits
of 1\,mJy with the VLA B array \citep{Sridharan02}.  This might
suggest contamination of the Arecibo beam by diffuse continuum
emission; as noted in \S\ref{subsec:velocity}, V32 is projected
against the bright patchy Sh2-86 emission.  On the other hand, V18 is
coincident with a small (20\arcsec) cometary optical nebula and the
Arecibo flux of 8\,mJy might be the more relevant if the VLA resolved
the extended emission. V18 may be another example of a HC \ion{H}{2}
region, and it is marked as a candidate in Figure~\ref{fig:lumvsmass}.
For V08 and V52 there are no sufficiently sensitive observations to
isolate any weak emission, or alternatively, to qualify them as HMPOs.

Most of the radio sources in this survey region are not in fact
associated with BLAST and {\it IRAS} sources (and vice versa).  One
must therefore be cautious of chance coincidences \citep[for reference
the surface density at the $\sim$1\,mJy level is $<
0.007$\,arcmin$^{-1}$,][]{fomalont1991}. There is a VGPS/NVSS source
offset by 75\arcsec\ from V40 and its {\it IRAS} counterpart, also
detected faintly at Arecibo \citep{Watson03}.  The level of radio
emission is at least an order of magnitude greater than would be
expected for the relatively low luminosity B2 star
\citep{panagia1973}, so that it is probably not associated at all.
Using the VLA B array at 8.2\,GHz, \citet{Sridharan02} placed a more
definitive upper limit of 1\,mJy.  Note that such a HMPO will never
become a significant \ion{H}{2} region unless the stellar mass is
increased through further accretion.
There are NVSS sources near a few other even lower luminosity BLAST
objects, V35, V39, V45, and V49 (and a marginal detection near V34).
They are not detected in the shallower MG surveys and so no radio
spectral index could be estimated.
However, V35 has been detected in the Westerbork Synthesis Radio
Telescope (WSRT) survey of \citet{taylor1996}. Its spectral index
between 327\,MHz and 1.4\,GHz is $-1.6$, and is therefore very
unlikely to be associated with thermal emission.  All of these sources
are seen projected on Sh2-86, and so the radio ``sources'' could be
just parts of the patchy ionization structure therein.  As mentioned
in \S\ref{sec:dist}, V39 is embedded at the end of a pillar
adjacent to an ionization front caused by the exciting stars of
NGC~6823, not V39.  Inspection of the radio images suggests that V45
and V49 might be similarly confused.

Clearly, more sensitive radio observations, with spatial and spectral
index information, are desirable to further elucidate the true nature
of all of these (luminous) sources.
But a firm conclusion that should be emphasized is that none of these
BLAST sources associated with NGC~6823 reveal a protostar or HMPO more
massive than a B0 to B1 star or about 20\,M$_\odot$.  There is thus no
evidence for a new (or lagging) generation of stars that are as
massive as the existing powerhouses of the NGC~6823 cluster, which
reach about 40\,M$_\odot$ \citep[O7~V and O9.5~Ia;][]{schaerer1997}.

\subsection{Extended Submillimeter Emission}
\label{sec:extended}

In this paper we have focused on the compact sources in the BLAST05
maps of Vulpecula. More diffuse structure could also be studied using
algorithms such as CLUMPFIND \citep{williams1994}. We leave such
quantitative analysis for future work. However, it is worth
illustrating a few features in the context of the discussion from
\S\ref{subsec:morphology} and the cool low luminosity-to-mass ratio
clumps already identified.

Between sources V02 and V03 there is a pair of extended BLAST
features, neither sufficiently compact to be included in our source
catalog. 
The first, to the west of the line, is a strong 100\micron\ source,
and even more prominent at 60\micron; it is {\it IRAS}~19364+2252.
This region exhibits a striking fan-shaped nebulosity in the IRAC
8\,\micron\ map (Figure~\ref{fig:iracblast}).  Since the object is
also a radio source, it is probably an evolved object. It has weak
emission in the BLAST maps and is anti-coincident with $^{13}$CO.
The second source, right on the line, is possibly related to the {\it
  IRAS} source 19367+2251, which IGA/MIGA show to be much weaker as a
compact source than the first.  This extended BLAST source coincides
with a peak in $^{13}$CO and becomes brighter relative to the first
source at increasing BLAST wavelengths.  This behavior implies that it
is a very cold object.

Another example is associated with the pair of {\it IRAS} sources
19385+2245 and 19389+2233.  They can be seen in all of the {\it IRAS}
bands, but they are both diffuse in the BLAST maps.  The first is weak
in $^{13}$CO and has complex structure in the 8\,\micron\ map. The
second is elongated in BLAST and has a strong elongated CO cloud as
well.  It is also extended at 8\,\micron, with a slight offset from the
submillimeter source.

A future analysis of such cold clouds in BLAST maps in conjunction
with MIPS and IRAC source catalogs may be used to determine whether
they represent an even earlier phase before condensations form, or if
they are simply distant cold envelopes of established young stellar
objects.

\section{CONCLUSIONS}  \label{sec:concl}

In this paper we present the first maps of the Galactic Plane observed
with BLAST during its 2005 LDB flight from Sweden to Canada,
specifically the survey in the direction of star forming clouds in
Vulpecula.

Fourier deconvolution is used to improve the angular resolution of the
maps from $\sim3\farcm5$ to $\sim1$\arcmin\ full-width half-power (the
theoretical 500\,\micron\ diffraction limit). In these maps, 60
compact submillimeter sources are detected simultaneously at 250, 350,
and 500\,\micron.  Complementary {\it IRAS}, MIPSGAL, and {\it MSX}
photometry are used to constrain the submillimeter-MIR SEDs of these
objects, and hence infer their cold dust temperatures and bolometric
fluxes. Our sample has convincingly revealed objects with a range of
dust temperatures, from $\sim12$--40\,K under the assumption of an
isothermal modified blackbody with dust emissivity index $\beta=1.5$.
With these SED fits, we derive luminosity-to-mass ratios (independent
of distance) in the range $0.2$--130\,L$_\odot$\,M$_\odot^{-1}$.
Those with low values are highly shielded and cool, with no evidence
for star formation, while those with high values have embedded
high-mass star formation.

Using a $^{13}$CO$(1 \rightarrow 0)$ data cube, the VLA Galactic Plane
Survey \ion{H}{1} cube, {\it Spitzer} IRAC and optical DSS images, we
argue that 49 of the 60 sources lie in a molecular cloud complex
associated with the open cluster NGC~6823 at $\sim2.3$\,kpc, 10
objects are associated with the Perseus arm at $\sim8.5$\,kpc and 1
object is in the outer Galaxy at $\sim14$\,kpc.  With these distance
estimates we calculate bolometric luminosities and cloud masses
associated with the thermal emission from cold dust. The most luminous
object ($7\times10^4$\,L$_\odot$) is in the Perseus arm.  Near
NGC~6823, the ranges are $\sim 40$--$10^4$\,L$_\odot$, and $\sim
15$--$700$\,M$_\odot$.

A mass function is constructed for the 49 objects associated with
NGC~6823. It is compatible with the spectrum of molecular gas masses
in other high-mass star forming regions, with a power-law index of
$-1.7$ for the $\ga100$\,M$_\odot$ sources. A flattening at lower
masses might be present, but is affected by detection completeness of
cool sources in this mass range.

The luminosity-mass distribution we find is broadly consistent with an
evolutionary sequence, from cool high-mass, low-luminosity clumps
(most are not detected with {\it IRAS} or {\it MSX} and there is no
evidence for any star formation), to more evolved HMPOs and UC
\ion{H}{2} regions.  But interestingly, among the embedded objects in
this molecular complex near NGC~6823 -- the next generation -- there
are none quite as massive as the 40\,M$_\odot$ stars currently
powering the nebula.

\acknowledgments

The BLAST collaboration acknowledges the support of NASA through grant
numbers NAG5-12785, NAG5-13301, and NNGO-6GI11G, the Canadian Space
Agency (CSA), the UK Particle Physics \& Astronomy Research Council
(PPARC), and Canada's Natural Sciences and Engineering Research
Council (NSERC). We would also like to thank the Columbia Scientific
Balloon Facility (CSBF) staff for their outstanding work.  We thank
Andy Gibb for helpful discussions. We also thank the anonymous referee
for useful comments. LO acknowledges partial support by the Puerto
Rico Space Grant Consortium and by the Fondo Institucional para la
Investigacion of the University of Puerto Rico, and also thanks
students Carlos M.  Poventud and Jorge L.  Morales for assistance with
the analysis. CBN acknowledges support from the Canadian Institute for
Advanced Research. DHH acknowledges the support of Consejo Nacional de
Ciencia y Technolog\'ia (CONACYt) grant 39953-F. This research has
been enabled by the use of WestGrid computing resources.

\bibliographystyle{apj}
\bibliography{apj-jour,refs}

\begin{deluxetable}{ccrrrrrrr}
\tablewidth{0pt} 
\small 
\tablecaption{BLAST Sources \label{tab:src}} 
\tablehead{ 
\colhead{BLAST} & 
\colhead{Source name} &
\colhead{$F_{250}$} & 
\colhead{$\sigma_{250}$} & 
\colhead{$F_{350}$} &
\colhead{$\sigma_{350}$} & 
\colhead{$F_{500}$} &
\colhead{$\sigma_{500}$} & 
\colhead{$V_{\mathrm{lsr}}$} \\
\colhead{ID} & 
\colhead{} & 
\colhead{(Jy)} & 
\colhead{(Jy)} &
\colhead{(Jy)} & 
\colhead{(Jy)} & 
\colhead{(Jy)} & 
\colhead{(Jy)} &
\colhead{km s$^{-1}$} } 
\startdata 
  V01 &BLAST J193837+230541 & 59.9 &  4.2 & 27.1 &  2.8 & 11.0 &  1.2 &  30.8\\
  V02 &BLAST J193849+230839 &258.0 &  5.1 & 97.8 &  3.4 & 36.9 &  1.3 &  32.3\\
  V03 &BLAST J193858+224637 &197.9 &  4.7 &101.9 &  3.5 & 44.5 &  1.3 &  36.5\\
  V04\tablenotemark{a} &BLAST J193921+224401 & 23.6 &  4.1 &  8.9 &  2.7 &  7.3 &  1.2 &  27.3\\
  V05 &BLAST J193935+235947 &325.7 &  5.6 &124.1 &  3.8 & 43.6 &  1.3 &  37.0\\
  V06 &BLAST J193943+232601 & 20.4 &  4.1 & 13.1 &  2.7 &  5.8 &  1.2 &  43.2\\
  V07\tablenotemark{b} &BLAST J194046+241007 & 38.7 &  4.1 & 12.2 &  2.7 &  5.2 &  1.1 & $-$53.5\\
  V08 &BLAST J194059+240439 &225.6 &  4.9 & 81.6 &  3.2 & 30.7 &  1.2 &  34.8\\
  V09 &BLAST J194103+240111 &194.8 &  4.7 &102.1 &  3.5 & 41.2 &  1.3 &  34.4\\
  V10 &BLAST J194106+235513 & 21.5 &  4.1 & 12.6 &  2.7 &  6.7 &  1.2 &  34.0\\
  V11 &BLAST J194136+232325 & 18.3 &  4.0 & 18.9 &  2.7 & 10.1 &  1.2 &  27.8\\
  V12 &BLAST J194144+231410 & 33.2 &  4.1 & 16.9 &  2.7 &  7.1 &  1.1 &  29.4\\
  V13 &BLAST J194144+231804 & 26.6 &  4.1 & 16.4 &  2.7 &  7.5 &  1.2 &  23.7\\
  V14 &BLAST J194146+234914 & 19.0 &  4.1 & 10.3 &  2.7 &  3.3 &  1.1 &  34.6\\
  V15 &BLAST J194213+232305 & 41.9 &  4.1 & 21.8 &  2.7 & 10.5 &  1.2 &  24.6\\
  V16 &BLAST J194217+231958 & 23.5 &  4.1 & 10.8 &  2.7 &  5.5 &  1.2 &  33.0\\
  V17\tablenotemark{c} &BLAST J194226+234532 & 30.4 &  4.1 & 10.8 &  2.7 &  5.6 &  1.2 &  $-$2.1\\
  V18 &BLAST J194228+230458 &259.9 &  5.1 &114.3 &  3.6 & 42.8 &  1.3 &  26.6\\
  V19\tablenotemark{ac} &BLAST J194232+234913 & 15.8 &  4.1 &  5.6 &  2.7 &  2.2 &  1.1 &  -3.0\\
  V20\tablenotemark{c} &BLAST J194232+234253 & 27.6 &  4.1 & 17.9 &  2.7 &  9.3 &  1.2 &   0.8\\
  V21\tablenotemark{c} &BLAST J194234+234412 & 20.9 &  4.1 & 15.6 &  2.7 &  6.6 &  1.2 &  $-$0.3\\
  V22 &BLAST J194234+232321 & 60.9 &  4.1 & 34.8 &  2.8 & 15.1 &  1.2 &  30.5\\
  V23\tablenotemark{c} &BLAST J194236+235003 &129.6 &  4.4 & 45.8 &  2.9 & 15.7 &  1.2 &  -3.0\\
  V24\tablenotemark{c} &BLAST J194244+234021 & 62.5 &  4.2 & 37.1 &  2.8 & 12.3 &  1.2 &   0.3\\
  V25 &BLAST J194246+225536 & 30.1 &  4.1 & 17.7 &  2.7 & 10.3 &  1.2 &  26.9\\
  V26 &BLAST J194254+232408 & 73.2 &  4.2 & 42.4 &  2.8 & 18.8 &  1.2 &  31.4\\
  V27 &BLAST J194306+230125 & 19.8 &  4.1 & 14.1 &  2.7 &  7.0 &  1.2 &  27.0\\
  V28 &BLAST J194308+232457 & 21.0 &  4.1 & 15.0 &  2.7 &  6.5 &  1.2 &  33.1\\
  V29\tablenotemark{a} &BLAST J194311+232010 & 15.7 &  4.1 &  2.3 &  2.7 &  5.5 &  1.2 &  24.0\\
  V30 &BLAST J194311+234405 &852.7 & 11.0 &378.5 &  8.5 &153.4 &  2.7 &  22.8\\
  V31 &BLAST J194311+232237 & 13.3 &  4.1 & 19.1 &  2.7 &  2.3 &  1.1 &  33.4\\
  V32 &BLAST J194317+231352 &221.2 &  4.9 &110.7 &  3.6 & 43.3 &  1.3 &  29.3\\
  V33 &BLAST J194319+232639 & 17.5 &  4.0 & 17.7 &  2.7 &  6.7 &  1.2 &  22.5\\
  V34\tablenotemark{a} &BLAST J194319+233906 & 27.5 &  4.1 & 24.1 &  2.7 &  7.4 &  1.2 &  22.9\\
  V35 &BLAST J194323+231512 & 22.0 &  4.1 &  7.9 &  2.7 &  1.4 &  1.1 &  30.4\\
  V36\tablenotemark{a} &BLAST J194325+234503 & 86.6 &  4.2 & 68.3 &  3.1 & 18.9 &  1.2 &  21.3\\
  V37 &BLAST J194326+240618 & 24.8 &  4.1 &  9.3 &  2.7 &  3.5 &  1.1 &  20.7\\
  V38\tablenotemark{a} &BLAST J194327+235638 & 64.4 &  4.2 & 43.8 &  2.9 & 22.6 &  1.2 &  22.4\tablenotemark{d}\\
  V39 &BLAST J194328+232032 & 57.8 &  4.1 & 31.1 &  2.8 & 12.6 &  1.2 &  25.7\\
  V40 &BLAST J194329+234013 &189.7 &  4.7 & 86.9 &  3.3 & 35.4 &  1.3 &  20.5\\
  V41 &BLAST J194331+235728 & 90.8 &  4.2 & 42.7 &  2.8 & 16.3 &  1.2 &  22.8\\
  V42\tablenotemark{a} &BLAST J194333+231445 & 23.8 &  4.1 & 24.0 &  2.7 &  7.5 &  1.2 &  26.9\\
  V43\tablenotemark{a} &BLAST J194334+232737 & 41.8 &  4.1 & 24.7 &  2.8 &  3.6 &  1.1 &  26.9\\
  V44 &BLAST J194335+235332 & 28.2 &  4.1 & 14.7 &  2.7 &  5.5 &  1.2 &  21.4\\
  V45\tablenotemark{a} &BLAST J194338+231725 & 10.6 &  4.1 &  7.0 &  2.7 &  0.4 &  1.1 &  26.6\\
  V46 &BLAST J194340+235506 & 28.8 &  4.1 & 12.5 &  2.7 &  5.3 &  1.1 &  22.5\\
  V47 &BLAST J194342+232006 &129.9 &  4.4 & 70.2 &  3.1 & 33.8 &  1.3 &  26.7\\
  V48 &BLAST J194344+240005 & 23.5 &  4.1 & 11.2 &  2.7 &  3.3 &  1.1 &  22.2\\
  V49\tablenotemark{a} &BLAST J194346+232532 & 27.0 &  4.1 & 17.9 &  2.7 & 11.5 &  1.2 &  26.1\\
  V50 &BLAST J194347+233505 & 18.6 &  4.1 & 10.1 &  2.7 &  3.0 &  1.1 &  22.1\\
  V51\tablenotemark{c} &BLAST J194349+230840 & 22.4 &  4.1 & 16.1 &  2.7 &  6.8 &  1.2 &  -7.0\\
  V52 &BLAST J194350+232839 &486.4 &  7.1 &214.5 &  5.3 & 84.2 &  1.8 &  27.6\\
  V53 &BLAST J194350+231846 & 40.6 &  4.1 & 20.0 &  2.7 &  9.1 &  1.2 &  27.7\\
  V54\tablenotemark{c} &BLAST J194351+234625 & 21.8 &  4.1 &  7.0 &  2.7 &  2.4 &  1.1 &  3.2\tablenotemark{e}\\
  V55\tablenotemark{a} &BLAST J194402+232409 & 30.6 &  4.1 & 21.9 &  2.7 &  5.0 &  1.2 &  29.3\\
  V56 &BLAST J194404+234418 & 53.5 &  4.1 & 37.3 &  2.8 & 15.6 &  1.2 &  21.7\\
  V57 &BLAST J194405+232658 & 40.7 &  4.1 & 19.6 &  2.7 &  7.1 &  1.2 &  27.2\\
  V58\tablenotemark{c} &BLAST J194409+230437 & 15.4 &  4.1 & 10.7 &  2.7 &  5.2 &  1.2 &  -12.7\\
  V59\tablenotemark{c} &BLAST J194424+234239 & 23.8 &  4.1 & 12.9 &  2.7 &  4.1 &  1.1 &  -10.7\\
  V60 &BLAST J194430+233426 & 14.9 &  4.1 & 14.1 &  2.7 &  4.8 &  1.2 &  26.2\\
\enddata
\tablecomments{Flux densities for BLAST sources are quoted at
  precisely 250, 350, and 500\,\micron\ using SED fits to obtain
  color-corrections for the band-averaged flux densities
  (\S\ref{sec:coldsed}). The quoted statistical uncertainties
  are determined from Monte Carlo simulations
  (\S\ref{sec:montecarlo}), and do not include calibration
  uncertainties.}
\tablenotetext{a}{These sources are located on ripples in the
  deconvolved map and the BLAST colors are considered unreliable as a
  result.}  \tablenotetext{b}{V07 is believed to lie in the outer
  galaxy.}  \tablenotetext{c}{These sources are associated with a
  molecular cloud in the Perseus arm.}  \tablenotetext{d}{Also has a
  comparable component at $-$15\,km\,s$^{-1}$.}
\tablenotetext{e}{Also has a comparable component at
  23\,km\,s$^{-1}$.}
\end{deluxetable}

\begin{deluxetable}{llrrrrrr}
\tablewidth{0pt}
\small
\tablecaption{{\it IRAS\/} PSC Counterparts \label{tab:iras}}
\tablehead{
\colhead{BLAST} &
\colhead{IRAS} &
\colhead{$\Delta \alpha$\tablenotemark{a}} &
\colhead{$\Delta \delta$\tablenotemark{a}} &
\colhead{$F_{12}$} &
\colhead{$F_{25}$} &
\colhead{$F_{60}$} &
\colhead{$F_{100}$} 
\\
\colhead{ID} &
\colhead{ID} &
\colhead{(\arcsec)} &
\colhead{(\arcsec)} &
\colhead{(Jy)} &
\colhead{(Jy)} &
\colhead{(Jy)} &
\colhead{(Jy)} 
}
\startdata
  V01 & 19364+2258 &$-$19.4 & $-$3.3 &    0.8 &    8.3 &  389.0\tablenotemark{b} &  698.0\tablenotemark{b}\\
  V02 & 19366+2301 &$-$21.0 & $-$1.0 &    8.8 &   62.0 &  389.0 &  698.0\\
  V03 & 19368+2239 & $-$4.5 & $-$5.4 &    2.3 &   11.7 &  112.0 &  231.0\\
  V05 & 19374+2352 &$-$27.0 &  7.3 &    4.6 &   24.3 &  422.0 &  768.0\\
  V07 & 19386+2403 & $-$5.9 &  2.1 &    1.8 &    4.7 &   71.2 &   63.8\\
  V08 & 19388+2357 & $-$3.0 & $-$0.1 &    1.3 &   10.9 &  254.0 &  433.0\\
  V09 & 19389+2354 &  1.5 & $-$0.1 &    1.8 &    1.2\tablenotemark{b} &   58.2 &  433.0\tablenotemark{b}\\
  V14 & 19396+2342 &  3.1 &  9.6 &    1.5\tablenotemark{b} &    0.8\tablenotemark{b} &    3.1 &   16.7\\
  V18 & 19403+2258 &$-$21.0 & 13.9 &   16.2 &  118.0 &  562.0 &  640.0\\
  V20 & 19404+2335 & $-$6.0 &  0.2 &    1.9\tablenotemark{b} &    1.0 &    4.9\tablenotemark{b} &   53.0\tablenotemark{b}\\
  V23\tablenotemark{c} & 19404+2342 &$-$24.0 & $-$7.9 & 9.7 & 35.3 &
  439.0 & 980.0\\
  V24 & 19406+2333 &  0.0 &  8.1 &    0.7 &    5.1 &   35.0 &   89.8\\
  V26 & 19407+2316 & $-$4.5 & $-$6.8 &    4.1\tablenotemark{b} &    4.2 &   23.2 &  165.0\tablenotemark{b}\\
  V30 & 19410+2336 &  4.5 &  0.9 &   14.4 &  109.0 &  982.0 & 1630.0\\
  V32 & 19411+2306 & 12.0 &  6.2 &    5.3 &   41.2 &  159.0 &  557.0\\
  V37 & 19413+2358 & $-$3.0 & $-$8.7 &   15.4\tablenotemark{b} &    0.8\tablenotemark{b} &    4.5 &   37.0\tablenotemark{b}\\
  V38\tablenotemark{d} & 19413+2349 & 10.5 & 20.1 &    0.7 &    4.6 &   42.5 &  132.0\\
  V40 & 19413+2332 & $-$4.5 & $-$9.3 &    4.2 &   38.5 &  217.0 & 1630.0\tablenotemark{b}\\
  V51 & 19416+2301 &  0.0 &$-$12.9 &    1.0\tablenotemark{b} &    1.0 &    4.7\tablenotemark{b} &   35.8\tablenotemark{b}\\
  V52 & 19416+2321 &$-$18.0 & 20.5 &    5.0\tablenotemark{b} &   24.7\tablenotemark{b} &  268.0 &  624.0\\
  V53\tablenotemark{d} & 19416+2312 &$-$39.0 & 39.3 &    0.6 &   10.7\tablenotemark{b} &   13.6\tablenotemark{b} &  250.0\\
  V57 & 19419+2319 &  4.5 &$-$15.0 &    2.8 &    2.6 &   36.1\tablenotemark{b} &  624.0\tablenotemark{b}\\
  V59 & 19422+2335 & $-$1.5 &  2.1 &    1.4 &    2.3 &   25.6 &   46.6\\
\enddata
\tablecomments{{\it IRAS} point sources associated
  with BLAST objects. The search radius is a variable function of both
  BLAST and {\it IRAS} positional uncertainties (see
  \S\ref{sec:iras}).}
\tablenotetext{a}{BLAST source tangent plane offsets (E and N)
  compared to {\it IRAS} source position.}
\tablenotetext{b}{These {\it IRAS} measurements are flagged as
  upper-limits in the catalog (usually due to confusion). Values from
  Table~\ref{tab:mapphot} are used instead.}
\tablenotetext{c}{IRAS 19404+2342 is close to both V19 and V23; the
  ambiguity is resolved with the improved resolution of the MIPS
  70\,\micron\ map.}
\tablenotetext{d}{Although IRAS 19413+2349 and 19416+2312 are close to
  V38 and V53 respectively, their SEDs are not compatible with the
  BLAST photometry and the identifications are not used in subsequent
  analyses.}
\end{deluxetable}

\begin{deluxetable}{lrcrcrcrcrc}
\tablewidth{0pt}
\small
\tablecaption{{\it IRAS} and {\it Spitzer} MIPS Photometry 
\label{tab:mapphot}}
\tablehead{
\colhead{BLAST} &
\colhead{$F_{12}$ } & 
\colhead{Flag} &
\colhead{$F_{25}$} &
\colhead{Flag} &
\colhead{$F_{60}$} &
\colhead{Flag} &
\colhead{$F_{100}$} &
\colhead{Flag} &
\colhead{$F_{70}$} &
\colhead{Flag} \\
\colhead{ID} &
\colhead{(Jy)} &
\colhead{} &
\colhead{(Jy)} &
\colhead{} &
\colhead{(Jy)} &
\colhead{} &
\colhead{(Jy)} &
\colhead{} &
\colhead{(Jy)} &
\colhead{}
}
\startdata
V01 & 1.2\tablenotemark{a} & g & 7.2\tablenotemark{a} & g & 46.5 & g & 120.5 & u & 33.0  & s \\
V02 & 12.1\tablenotemark{a} & g & 60.9\tablenotemark{a} & g & 437.7\tablenotemark{a} & g & 733.6\tablenotemark{a} & g & 211.1  & s \\
V03 & 2.4\tablenotemark{a} & g & 12.4\tablenotemark{a} & g & 114.5\tablenotemark{a} & g & 243.2\tablenotemark{a} & g & 78.2  & s \\
V04 & 0.2 & n & 0.2 & n & 0.2 & n & 31.3 & n & 0.4  & n \\
V05 & 5.9\tablenotemark{a} & g & 28.6\tablenotemark{a} & g & 410.5\tablenotemark{a} & g & 772.9\tablenotemark{a} & g & 218.6  & s \\
V06 & $-$0.1 & n & $-$0.1 & n & 4.2 & g & 24.7 & u & 2.5  & g \\
V07 & 1.7\tablenotemark{a} & g & 4.7\tablenotemark{a} & g & 66.7\tablenotemark{a} & g & 135.7\tablenotemark{a} & g & 50.4  & s \\
V08 & 2.3\tablenotemark{a} & g & 12.1\tablenotemark{a} & g & 259.9\tablenotemark{a} & g & 569.6\tablenotemark{a} & g & 206.7  & s \\
V09 & 1.5\tablenotemark{a} & g & 2.9 & g & 44.1\tablenotemark{a} & g & 256.8 & u & 44.1  & s \\
V10 & 0.0 & n & 0.0 & g & 2.4 & u & 48.3 & u & 0.9  & n \\
V11 & $-$0.1 & n & $-$0.2 & n & $-$0.2 & n & 18.2 & n & 0.8  & n \\
V12 & 1.7 & g & 4.0 & g & 14.8 & g & 70.9 & u & 9.4  & g \\
V13 & $-$0.4 & n & $-$0.4 & n & $-$3.6 & n & 13.0 & n & $-$1.5  & n \\
V14 & 0.3 & n & 0.5 & u & 4.0\tablenotemark{a} & g & 28.0\tablenotemark{a} & u & 2.7  & g \\
V15 & $-$0.2 & n & 0.2 & n & 9.4 & u & 83.3 & u & 3.1  & n \\
V16 & 1.2 & u & 2.0 & n & 11.7 & u & 91.9 & u & 3.5  & g \\
V17 & 0.2 & n & 0.9 & u & 5.9 & u & 14.5 & n & 3.4  & n \\
V18 & 17.2\tablenotemark{a} & g & 108.5\tablenotemark{a} & g & 570.0\tablenotemark{a} & g & 699.1\tablenotemark{a} & g & 275.4  & s \\
V19 & 17.6 & u & 52.8 & u & 460.1 & u & 979.5 & u & 148.6  & s \\
V20 & 0.0 & g & 0.3\tablenotemark{a} & g & 6.6 & u & 19.4 & n & 6.3  & g \\
V21 & $-$0.1 & n & $-$0.3 & n & $-$1.9 & n & $-$0.1 & n & 4.5  & g \\
V22 & 0.7 & n & 0.5 & n & 7.4 & u & 97.7 & u & 6.0  & n \\
V23 & 15.3\tablenotemark{a} & u & 53.3\tablenotemark{a} & u & 506.5\tablenotemark{a} & g & 916.8\tablenotemark{a} & g & 192.2  & s \\
V24 & 1.4\tablenotemark{a} & u & 3.5\tablenotemark{a} & g & 32.0\tablenotemark{a} & g & 110.6\tablenotemark{a} & u & 25.5  & s \\
V25 & 0.2 & n & 0.9 & u & 3.8 & g & 47.0 & g & 0.7  & g \\
V26 & 2.0 & u & 4.0\tablenotemark{a} & g & 21.9\tablenotemark{a} & g & 91.4 & u & 7.7  & g \\
V27 & 0.4 & u & 0.7 & u & 7.7 & u & 51.7 & u & 0.7  & n \\
V28 & 2.6 & u & 4.2 & u & $-$3.3 & n & 36.9 & n & 3.2  & n \\
V29 & $-$0.7 & n & 0.1 & n & $-$0.5 & n & $-$6.2 & n & $-$0.9  & g \\
V30 & 18.5\tablenotemark{a} & g & 118.0\tablenotemark{a} & g & 892.2\tablenotemark{a} & g & 1463.9\tablenotemark{a} & g & 378.2  & s \\
V31 & 0.7 & n & 0.9 & u & 2.0 & n & $-$7.8 & n & $-$0.3  & n \\
V32 & 6.1\tablenotemark{a} & g & 37.3\tablenotemark{a} & g & 151.2\tablenotemark{a} & g & 478.2\tablenotemark{a} & g & 62.6  & s \\
V33 & 1.2 & n & 1.9 & u & 21.5 & u & 37.0 & u & $-$1.3  & n \\
V34 & 3.2 & u & 5.4 & u & 83.9 & u & 334.1 & u & 2.2  & n \\
V35 & 6.2 & u & 22.9 & u & 87.9 & u & 461.9 & u & 6.6  & n \\
V36 & 5.6 & u & 8.6 & g & 88.4 & u & 472.2 & u & 42.6  & s \\
V37 & 0.3 & n & 0.8 & n & 4.4\tablenotemark{a} & g & 24.6 & u & 3.3  & g \\
V38 & 0.9 & u & 5.3 & u & 34.0 & u & 120.9 & u & 22.9  & s \\
V39 & 3.8 & u & 3.7 & g & 40.4 & u & 183.2 & u & 19.4  & g \\
V40 & 7.2\tablenotemark{a} & g & 44.4\tablenotemark{a} & g & 214.3\tablenotemark{a} & g & 400.5 & g & 152.3  & s \\
V41 & 0.6 & n & 5.2 & u & 33.5 & u & 131.7 & u & 17.8  & u \\
V42 & $-$0.8 & n & $-$0.6 & n & $-$0.6 & n & 223.5 & u & $-$3.3  & n \\
V43 & 1.8 & n & 1.6 & n & 21.1 & u & 147.6 & u & 7.8  & n \\
V44 & 0.0 & n & 0.1 & n & 3.6 & u & 35.2 & n & 2.0  & u \\
V45 & 0.0 & n & 3.9 & u & 1.7 & n & 17.7 & n & 6.8  & n \\
V46 & 0.0 & n & $-$0.2 & n & 9.5 & u & 137.3 & u & 12.3  & u \\
V47 & 0.9 & g & 10.5 & g & 60.5 & g & 174.6 & g & 21.0  & s \\
V48 & 0.8 & n & 1.2 & n & 9.0 & u & 61.0 & u & 8.2  & g \\
V49 & $-$0.9 & n & 1.0 & n & 16.7 & u & 149.0 & u & 17.6  & g \\
V50 & 0.2 & g & 1.0 & g & 6.7 & u & 51.5 & u & 4.2  & g \\
V51 & 0.3 & n & 1.9\tablenotemark{a} & n & 15.9 & u & 103.5 & u & 3.2  & g \\
V52 & 4.4 & u & 27.6 & u & 278.2\tablenotemark{a} & g & 675.1\tablenotemark{a} & g & 211.7  & s \\
V53 & $-$0.5\tablenotemark{a} & n & $-$0.3 & n & 5.5 & n & 227.9\tablenotemark{a} & u & 0.7  & n \\
V54 & 0.5 & u & 0.9 & u & 11.2 & u & 82.4 & u & 12.9  & g \\
V55 & 1.9 & u & 5.6 & u & 11.3 & g & 369.6 & u & 8.1  & n \\
V56 & $-$0.1 & n & $-$0.3 & n & 1.1 & n & 31.4 & u & $-$0.4  & n \\
V57 & 1.7\tablenotemark{a} & g & 5.3\tablenotemark{a} & g & 4.5 & n & 423.2 & u & 5.7  & g \\
V58 & $-$0.1 & n & $-$0.2 & n & 7.2 & u & 90.1 & u & 2.5  & g \\
V59 & 1.4\tablenotemark{a} & g & 2.3\tablenotemark{a} & g & 26.4\tablenotemark{a} & g & 57.4\tablenotemark{a} & g & 24.7  & s \\
V60 & 0.8 & n & 1.3 & u & 10.6 & u & 75.4 & u & 2.2  & n \\

\enddata
\tablecomments{{\it IRAS} and {\it Spitzer} MIPS map photometry are
  described in \S\ref{sec:iras} and \S\ref{sec:mips70}
  respectively.  Each measurement is flagged upon visual inspection:
  ``g'' if it is good; ``u'' for upper limit; ``n'' for no visible
  source; and ``s'' for saturated (unreliable) values in the MIPS
  70\,\micron\ map.}
\tablenotetext{a}{Clear detections from the {\it IRAS} PSC
  (Table~\ref{tab:iras}) are used in favor of these values when
  available.}
\end{deluxetable}

\begin{deluxetable}{lrrrrrr}
\tablewidth{0pt}
\small
\tablecaption{MSX counterparts \label{tab:msx}}
\tablehead{
\colhead{BLAST} &
\colhead{$\Delta \alpha$\tablenotemark{a}} &
\colhead{$\Delta \delta$\tablenotemark{a}} &
\colhead{$F_8.3$} &
\colhead{$F_{12.1}$} &
\colhead{$F_{14.7}$} &
\colhead{$F_{21.3}$} \\
\colhead{ID} &
\colhead{(\arcsec)} &
\colhead{(\arcsec)} &
\colhead{(Jy)} &
\colhead{(Jy)} &
\colhead{(Jy)} &
\colhead{(Jy)}
}
\startdata
V01 & $-$6.9 & 1.4 &   0.9 &   0.8 &   0.8 &   4.7\\
V02 & $-$11.0 & $-$1.3 &   4.3 &   5.3 &   3.7 &  23.9\\
V03 & 18.0 & 8.1 &   0.9 &   1.1 &   0.6 &   1.6\\
 & $-$20.7 & $-$9.9 &   0.8 &   1.0 &   1.4 &   3.4\\
V05 & $-$6.9 & 26.0 &   1.7 &   1.8 &   1.2 &   2.7\\
 & $-$2.7 & 2.9 &   0.9 &   0.8 &   1.2 &   6.2\\
V06 & 19.3 & $-$63.1 &   0.4 & \nodata & \nodata & \nodata\\
V07 & $-$0.0 & 11.2 &   1.0 &   1.4 &   0.8 &   2.0\\
 & $-$26.0 & $-$53.6 &   0.1 &   1.2 & \nodata & \nodata\\
V08 & $-$0.0 & $-$23.9 &   0.6 &   1.4 &   0.9 & \nodata\\
 & $-$8.2 & 6.7 &   0.5 &   1.3 &   0.8 &   3.6\\
V12 & $-$48.2 & $-$41.7 &   1.7 &   1.2 &   0.7 & \nodata\\
 & 8.3 & 9.8 &   0.5 &   2.0 &   3.1 &   4.7\\
V13 & $-$4.1 & 21.1 &   0.1 & \nodata & \nodata & \nodata\\
V14 & $-$11.0 & 16.7 &   0.2 & \nodata &   0.8 & \nodata\\
V18 & $-$8.3 & 16.9 &   8.4 &   9.0 &   6.2 &  57.5\\
V19 & 11.0 & 28.5 &   0.9 &   1.8 &   2.2 &   3.9\\
 & 37.0 & 38.2 &   0.7 &   2.6 &   3.2 &   7.9\\
 & $-$9.6 & 49.0 &   0.1 &   0.6 & \nodata &   1.3\\
V20 & $-$0.0 & $-$0.2 &   0.2 & \nodata & \nodata & \nodata\\
V21 & $-$15.1 & 13.9 &   0.1 & \nodata & \nodata & \nodata\\
V22 & $-$20.7 & $-$3.6 &   0.2 & \nodata & \nodata & \nodata\\
V23 & $-$15.1 & $-$11.7 &   0.7 &   2.6 &   3.2 &   7.9\\
 & 17.8 & $-$5.2 &   1.5 &   2.5 &   1.3 &   3.1\\
 & $-$19.2 & 17.1 &   0.8 &   2.2 &   2.7 &   5.2\\
V24 & $-$5.5 & 9.9 &   0.4 & \nodata & \nodata &   2.5\\
V26 & 15.1 & 6.7 &   1.4 &   2.1 &   2.2 &   3.6\\
V28 & $-$23.4 & $-$28.5 &   0.9 &   1.4 &   1.7 &   1.9\\
V30 & $-$0.0 & $-$0.2 &   5.3 &   8.5 &  14.9 &  47.4\\
V31 & $-$15.1 & 17.8 &   0.2 & \nodata & \nodata & \nodata\\
V32 & 8.3 & 8.5 &   1.8 &   5.2 &   8.7 &  25.6\\
V35 & 66.1 & 25.8 &   0.2 & \nodata & \nodata &   2.7\\
V36 & 31.6 & $-$15.4 &   1.7 &   1.6 &   1.0 &   2.1\\
V37 & 20.5 & 60.9 &   0.9 &   0.6 &   0.5 & \nodata\\
V38 & 2.7 & 16.0 &   0.7 &   0.9 &   1.2 &   2.7\\
V39 & 2.8 & 10.5 &   0.9 &   1.9 &   2.1 &   2.3\\
V40 & $-$9.6 & $-$21.8 &   2.2 &   2.5 &   1.1 &   6.1\\
 & 6.9 & 11.7 &   2.0 &   2.2 &   2.7 &  18.3\\
V43 & $-$6.9 & $-$20.0 &   0.2 & \nodata & \nodata & \nodata\\
V46 & 8.2 & 4.9 &   0.1 & \nodata & \nodata & \nodata\\
 & 4.1 & $-$39.0 &   0.3 &   1.8 & \nodata & \nodata\\
V47 & 4.1 & 8.8 &   0.1 &   1.1 &   1.7 &   4.7\\
V49 & $-$19.3 & $-$24.1 &   0.4 &   0.8 & \nodata & \nodata\\
V50 & 15.1 & 48.0 &   0.2 & \nodata & \nodata & \nodata\\
V51 & $-$2.8 & $-$7.9 &   0.2 & \nodata & \nodata & \nodata\\
V52 & $-$4.1 & 0.0 &   0.2 & \nodata & \nodata &   3.3\\
V53 & $-$20.7 & $-$29.8 &   0.1 & \nodata & \nodata & \nodata\\
V55 & $-$17.9 & 11.8 &   0.2 & \nodata & \nodata & \nodata\\
V57 & $-$1.4 & $-$11.7 &   1.3 &   1.3 &   1.2 &   1.2\\
V59 & 1.4 & 1.5 &   0.8 &   1.1 &   0.6 & \nodata\\
V60 & $-$52.2 & 5.7 &   0.1 &   1.0 & \nodata & \nodata\\

\enddata
\tablecomments{All candidate {\it MSX} counterparts are listed for
each BLAST source (\S\ref{sec:msx}).}
\tablenotetext{a}{BLAST source tangent plane offsets (E and N)
  compared to {\it MSX} source position.}
\end{deluxetable}

\begin{deluxetable}{lcrrcc}
\tablewidth{0pt}
\small
\tablecaption{Results from SED Fits \label{tab:sed}}
\tablehead{
\colhead{BLAST} &
\colhead{Temperature} &
\colhead{$M_\mathrm{c}$} &
\colhead{$L_{\mathrm{FIR}}$} &
\colhead{Luminosity Fraction} &
\colhead{$L_{\mathrm{FIR}}/M_\mathrm{c}$} \\
\colhead{ID} &
\colhead{(K)} &
\colhead{(M$_\odot$)} &
\colhead{(L$_\odot$)} &
\colhead{(\%)} &
\colhead{(L$_\odot$\,M$_\odot^{-1}$)}
}
\startdata
V01 & $ 27.2 \pm  1.5 $ & $ 44 \pm 7 $ & $ 750 \pm 130 $ & $ 76 $ &  $   17.5 \pm   5.0 $ \\
V02 & $ 31.9 \pm  2.3 $ & $ 114 \pm 22 $ & $ 4500 \pm 1000 $ & $ 78 $ &  $   40.5 \pm  15.8 $ \\
V03 & $ 22.2 \pm  0.9 $ & $ 273 \pm 40 $ & $ 1470 \pm 150 $ & $ 67 $ &  $    5.6 \pm   1.2 $ \\
V04 & $ 20.0 $ & $ 38 \pm 5 $ & $ 121 \pm 15 $ & \nodata &  $    3.2 $ \\
V05 & $ 31.3 \pm  1.9 $ & $ 130 \pm 27 $ & $ 4910 \pm 880 $ & $ 89 $ &  $   36.1 \pm  12.2 $ \\
V06 & $ 20.2 \pm  2.4 $ & $ 55 \pm 35 $ & $ 96 \pm 57 $ & $ 62 $ &  $    1.9 \pm   1.9 $ \\
V07 & $ 32.6 \pm  1.9 $ & $ 579 \pm 122 $ & $ 28200 \pm 4900 $ & $ 83 $ &  $   49.0 \pm  16.5 $ \\
V08 & $ 29.7 \pm  1.7 $ & $ 97 \pm 19 $ & $ 2610 \pm 430 $ & $ 84 $ &  $   26.5 \pm   8.8 $ \\
V09 & $ 22.8 \pm  1.7 $ & $ 233 \pm 45 $ & $ 1560 \pm 350 $ & $ 80 $ &  $    6.4 \pm   2.4 $ \\
V10 & $ 15.8 \pm  2.6 $ & $ 89 \pm 50 $ & $ 53 \pm 22 $ & $ 93 $ &  $    0.9 \pm   0.9 $ \\
V11 & $ 12.3 \pm  1.6 $ & $ 213 \pm 125 $ & $ 42 \pm 15 $ & \nodata &  $    0.2 \pm   0.1 $ \\
V12 & $ 21.9 \pm  1.0 $ & $ 42 \pm 8 $ & $ 220 \pm 30 $ & $ 51 $ &  $    5.3 \pm   1.3 $ \\
V13 & $ 16.5 \pm  1.5 $ & $ 85 \pm 28 $ & $ 87 \pm 20 $ & $ 70 $ &  $    1.0 \pm   0.5 $ \\
V14 & $ 22.6 \pm  1.0 $ & $ 17 \pm 5 $ & $ 107 \pm 11 $ & $ 70 $ &  $    6.0 \pm   1.6 $ \\
V15 & $ 17.8 \pm  1.5 $ & $ 93 \pm 23 $ & $ 145 \pm 43 $ & \nodata &  $    1.5 \pm   0.7 $ \\
V16 & $ 20.2 \pm  2.0 $ & $ 39 \pm 16 $ & $ 117 \pm 34 $ & \nodata &  $    2.9 \pm   1.6 $ \\
V17 & $ 19.0 \pm  1.1 $ & $ 651 \pm 171 $ & $ 1400 \pm 280 $ & \nodata &  $    2.3 \pm   0.8 $ \\
V18 & $ 30.3 \pm  1.5 $ & $ 148 \pm 24 $ & $ 4590 \pm 740 $ & $ 58 $ &  $   30.9 \pm   9.3 $ \\
V19 & $ 20.0 $ & $ 281 \pm 67 $ & $ 890 \pm 210 $ & \nodata &  $    3.2 $ \\
V20 & $ 16.6 \pm  3.1 $ & $ 1274 \pm 739 $ & $ 1250 \pm 590 $ & $ 75 $ &  $    0.9 \pm   0.9 $ \\
V21 & $ 15.8 \pm  3.2 $ & $ 1109 \pm 762 $ & $ 920 \pm 540 $ & $ 66 $ &  $    0.9 \pm   0.9 $ \\
V22 & $ 17.7 \pm  1.3 $ & $ 145 \pm 34 $ & $ 227 \pm 44 $ & $ 72 $ &  $    1.6 \pm   0.6 $ \\
V23 & $ 39.4 \pm  1.7 $ & $ 526 \pm 65 $ & $ 71000 \pm 11000 $ & $ 89 $ &  $  130.4 \pm  31.6 $ \\
V24 & $ 23.2 \pm  0.8 $ & $ 970 \pm 121 $ & $ 6770 \pm 640 $ & $ 68 $ &  $    6.7 \pm   1.0 $ \\
V25 & $ 17.6 \pm  2.6 $ & $ 93 \pm 44 $ & $ 112 \pm 61 $ & $100 $ &  $    1.3 \pm   1.1 $ \\
V26 & $ 18.9 \pm  1.1 $ & $ 150 \pm 27 $ & $ 346 \pm 62 $ & $ 58 $ &  $    2.4 \pm   0.7 $ \\
V27 & $ 14.9 \pm  2.6 $ & $ 105 \pm 64 $ & $ 54 \pm 26 $ & \nodata &  $    0.6 \pm   0.6 $ \\
V28 & $ 15.9 \pm  1.8 $ & $ 90 \pm 46 $ & $ 60 \pm 25 $ & $ 39 $ &  $    0.8 \pm   0.7 $ \\
V29 & $ 20.0 $ & $ 23 \pm 4 $ & $ 73 \pm 14 $ & \nodata &  $    3.2 $ \\
V30 & $ 25.9 \pm  1.0 $ & $ 680 \pm 115 $ & $ 9020 \pm 700 $ & $ 63 $ &  $   13.0 \pm   2.9 $ \\
V31 & $ 16.8 \pm  1.7 $ & $ 43 \pm 14 $ & $ 45 \pm 22 $ & $ 50 $ &  $    1.0 \pm   0.6 $ \\
V32 & $ 26.2 \pm  1.3 $ & $ 208 \pm 30 $ & $ 2820 \pm 410 $ & $ 69 $ &  $   13.7 \pm   3.7 $ \\
V33 & $ 14.2 \pm  2.3 $ & $ 107 \pm 64 $ & $ 49 \pm 21 $ & \nodata &  $    0.4 \pm   0.4 $ \\
V34 & $ 20.0 $ & $ 52 \pm 6 $ & $ 166 \pm 18 $ & \nodata &  $    3.2 $ \\
V35 & $ 20.7 \pm  1.2 $ & $ 22 \pm 7 $ & $ 84 \pm 19 $ & $ 54 $ &  $    4.0 \pm   1.1 $ \\
V36 & $ 20.0 $ & $ 118 \pm 12 $ & $ 374 \pm 39 $ & $ 54 $ &  $    3.2 $ \\
V37 & $ 22.6 \pm  0.9 $ & $ 24 \pm 6 $ & $ 139 \pm 18 $ & $ 73 $ &  $    5.8 \pm   1.5 $ \\
V38 & $ 20.0 $ & $ 132 \pm 13 $ & $ 417 \pm 42 $ & $ 72 $ &  $    3.2 $ \\
V39 & $ 21.7 \pm  0.6 $ & $ 78 \pm 8 $ & $ 390 \pm 32 $ & $ 70 $ &  $    5.0 \pm   0.7 $ \\
V40 & $ 28.0 \pm  1.6 $ & $ 142 \pm 26 $ & $ 2670 \pm 610 $ & $ 66 $ &  $   18.3 \pm   6.9 $ \\
V41 & $ 20.8 \pm  0.8 $ & $ 110 \pm 14 $ & $ 423 \pm 43 $ & \nodata &  $    3.8 \pm   0.8 $ \\
V42 & $ 20.0 $ & $ 49 \pm 6 $ & $ 155 \pm 18 $ & \nodata &  $    3.2 $ \\
V43 & $ 20.0 $ & $ 45 \pm 5 $ & $ 142 \pm 15 $ & $ 86 $ &  $    3.2 $ \\
V44 & $ 18.3 \pm  1.3 $ & $ 53 \pm 15 $ & $ 103 \pm 21 $ & \nodata &  $    2.0 \pm   0.7 $ \\
V45 & $ 20.0 $ & $ 14 \pm 4 $ & $ 46 \pm 13 $ & \nodata &  $    3.2 $ \\
V46 & $ 21.4 \pm  1.7 $ & $ 37 \pm 12 $ & $ 163 \pm 35 $ & $ 66 $ &  $    4.5 \pm   1.4 $ \\
V47 & $ 20.2 \pm  2.9 $ & $ 219 \pm 70 $ & $ 830 \pm 310 $ & $ 61 $ &  $    3.4 \pm   2.3 $ \\
V48 & $ 22.5 \pm  1.7 $ & $ 25 \pm 7 $ & $ 154 \pm 34 $ & \nodata &  $    5.8 \pm   2.3 $ \\
V49 & $ 20.0 $ & $ 64 \pm 6 $ & $ 203 \pm 18 $ & $ 58 $ &  $    3.2 $ \\
V50 & $ 21.1 \pm  2.1 $ & $ 25 \pm 10 $ & $ 101 \pm 31 $ & $ 72 $ &  $    4.0 \pm   2.4 $ \\
V51 & $ 16.6 \pm  3.4 $ & $ 1088 \pm 727 $ & $ 1050 \pm 570 $ & $ 70 $ &  $    1.2 \pm   1.2 $ \\
V52 & $ 24.6 \pm  1.4 $ & $ 390 \pm 75 $ & $ 3850 \pm 610 $ & $ 84 $ &  $    9.7 \pm   3.1 $ \\
V53 & $ 18.5 \pm  1.3 $ & $ 80 \pm 17 $ & $ 156 \pm 34 $ & $ 80 $ &  $    1.8 \pm   0.7 $ \\
V54 & $ 24.9 \pm  1.3 $ & $ 210 \pm 48 $ & $ 2320 \pm 410 $ & \nodata &  $   10.5 \pm   3.2 $ \\
V55 & $ 20.0 $ & $ 48 \pm 6 $ & $ 153 \pm 17 $ & $ 59 $ &  $    3.2 $ \\
V56 & $ 16.1 \pm  1.3 $ & $ 180 \pm 42 $ & $ 176 \pm 49 $ & \nodata &  $    0.9 \pm   0.5 $ \\
V57 & $ 20.1 \pm  1.1 $ & $ 58 \pm 9 $ & $ 181 \pm 31 $ & $ 60 $ &  $    3.1 \pm   1.0 $ \\
V58 & $ 16.7 \pm  3.9 $ & $ 757 \pm 510 $ & $ 540 \pm 330 $ & \nodata &  $    0.8 \pm   0.7 $ \\
V59 & $ 25.4 \pm  1.5 $ & $ 297 \pm 67 $ & $ 3400 \pm 390 $ & $ 63 $ &  $   12.2 \pm   4.4 $ \\
V60 & $ 15.8 \pm  2.3 $ & $ 65 \pm 28 $ & $ 42 \pm 21 $ & $ 34 $ &  $    0.6 \pm   0.5 $ \\
\enddata
\tablecomments{Results from modified blackbody SED
  (Equation~\ref{eq:sed}) fits to submillimeter and FIR photometry
  (\S\ref{sec:coldsed}). These fits assume a dust emissivity index
  $\beta=1.5$.  Cloud masses $M_\mathrm{c}$ are derived from the
  amplitude $A$ of the SED using Equation~\ref{eq:mass} and distances
  from \S\ref{sec:dist}.  FIR luminosities, $L_{\mathrm{FIR}}$,
  are the integrated luminosities from the modified blackbody. Objects
  with MIR photometry have estimates for the total bolometric
  luminosity, which may be obtained by dividing the FIR luminosity by
  the Luminosity Fraction. Quoted uncertainties are statistical errors
  derived from Monte Carlo simulations (\S\ref{sec:sedfits}).
  Other systematic errors are not included. For example, adopting
  $\beta=2.0$ decreases temperatures by $\sim5$\,K, and increases
  cloud masses by a factor $\sim2$ (\S\ref{sec:sedfits}). There
  are also significant distance uncertainties
  (\S\ref{sec:dist}), and a factor of $\sim2$ uncertainty
  introduced through choices of $\kappa_0$ and $R$ in
  Equation~\ref{eq:mass}. }
\end{deluxetable}

\end{document}